\documentclass[aps,prx,reprint,superscriptaddress,twocolumn,amsmath,amssymb,amsfonts]{revtex4-1}

\AtBeginDocument{\usepackage{booktabs}}               
\makeatletter
\g@addto@macro\bfseries{\boldmath}
\makeatother

\usepackage[varg]{txfonts}
\usepackage[T1]{fontenc}
\usepackage[utf8]{inputenc}
\usepackage{hyperref}
\usepackage{amsmath}
\usepackage{color}
\usepackage{graphicx}
\usepackage[percent]{overpic}
\usepackage{mathrsfs}
\usepackage{bm}
\usepackage{braket}
\usepackage[nolist,nohyperlinks]{acronym}
\usepackage{comment}
\usepackage{tikz}

\newcommand{\avg}[1]{\braket{#1}}

\newcommand{\h}[1]{{#1}^{\dagger}}
\newcommand{\f}[1]{{#1}^{}}
\newcommand{\cc}[1]{{#1}^{*}}

\renewcommand{\vec}[1]{\boldsymbol{#1}}
\newcommand{\mat}[1]{\vec{#1}}

\newcommand{\vhat}[1]{\vec{\hat{#1}}}

\newcommand{\meV}{\ {\rm meV}}
\newcommand{\T}{\ {\rm T}}
\newcommand{\s}{\sigma}

\definecolor{cred}{RGB}{228,26,28}
\definecolor{cblue}{RGB}{55,126,184}
\definecolor{cdblue}{RGB}{40,96,139}
\definecolor{clblue}{RGB}{205,223,237}
\definecolor{cgreen}{RGB}{77,175,74}
\definecolor{cgray}{RGB}{150,150,150}
\definecolor{clgray}{RGB}{200,200,200}
\definecolor{cpurple}{RGB}{152,78,163}
\definecolor{corange}{RGB}{255,127,0}
\definecolor{cgold}{RGB}{230,171,2}
\definecolor{cL}{RGB}{255,0,0}
\hypersetup{colorlinks=true,linkcolor=cL,citecolor=cL,urlcolor=cL} 

\begin{document}

\title{Magnetoelectric generation of a Majorana-Fermi surface in Kitaev's honeycomb model}
\author{Rajas Chari} 
\affiliation{Max-Planck-Institut f\"ur Physik komplexer Systeme, 01187 Dresden, Germany} 
\affiliation{Department of Physics, Indian Institute of Technology Madras, Chennai 600036, India} 
\author{Roderich Moessner} 
\affiliation{Max-Planck-Institut f\"ur Physik komplexer Systeme, 01187 Dresden, Germany}
\author{Jeffrey G. Rau} 
\affiliation{Max-Planck-Institut f\"ur Physik komplexer Systeme, 01187 Dresden, Germany}
\affiliation{Department of Physics, University of Windsor, Windsor, Ontario, N9B 3P4, Canada}

\begin{abstract}
We study the effects of static magnetic and electric fields on Kitaev's honeycomb model. Using the electric polarization operator appropriate for Kitaev materials, we derive the effective Hamiltonian for the emergent Majorana fermions to second-order in both the electric and magnetic fields. We find that while individually each perturbation does not qualitatively alter Kitaev spin liquid, the cross-term induces a finite chemical potential at each Dirac node, giving rise to a Majorana-Fermi surface. We argue this gapless phase is stable and exhibits typical metallic phenomenology, such as linear in temperature heat capacity and finite, but non-quantized, thermal Hall response. Finally, we speculate on the potential for realization of this physics in Kitaev materials.
\end{abstract}

\date{\today}

\maketitle

\section{Introduction}
\label{sec:intro}
Topological states of matter have attracted broad interest due to their fundamental importance in our understanding of many-body systems~\cite{savary2016quantum}, as well as their potential practical importance in storing and manipulating quantum information~\cite{pachos2012introduction}. A key role in our understanding has been played by exactly solvable models, such as the toric code~\cite{kitaev2003fault}, where the nature of the ground state and fractionalized excitations is indisputable. However, finding and exploring the physics of topological phases of matter, such as spin liquids, in more realistic models is difficult, with fewer tractable systems to study~\cite{knolle2019field}.

Kitaev's honeycomb model~\cite{kitaev2006} represents a rare example of a class of exactly solvable models of spin liquids that may, to a reasonable approximation, be realized in solid-state magnetic systems~\cite{rau2016spin,winter2017models,takagi2019concept}. Specifically, in transition metal magnets with strong spin-orbit coupling, a microscopic super-exchange mechanism has been identified for an edge-shared bonding geometry~\cite{jackeli2009mott} that yields Kitaev's anisotropic exchange interaction at leading order~\cite{chaloupka2010kitaev,rau2014,katukuri2014kitaev}.
As such, Kitaev's model has been the subject of intense study~\cite{hermanns2018physics} to determine its response to a variety of perturbations and probes, including mapping the nearby phase diagram~\cite{chaloupka2010kitaev,chaloupka2013zigzag,rau2014,katukuri2014kitaev,rau2014trigonal,chaloupka2015hidden} with an eye towards materials, understanding thermal properties~\cite{nasu2014vaporization,nasu2015thermal,nasu2017thermal,yoshitake2016fractional}, computing its dynamical responses~\cite{knolle2014dynamics,knolle2014raman,knolle2015dynamics,smith2015neutron,perreault2015theory,halasz2016resonant}, generalizing it to three-dimensional lattices~\cite{mandal2009,nasu2014finite,hermanns2014oct,hermanns2015weyl,hermanns2015spin,o2016classification,mishchenko2017finite}, as well as understanding the effects of disorder~\cite{willans2010disorder,willans2011site,zschocke2015physical,udagawa2018vison,otten2019dynamical,knolle2019bond}. 

More recently, the effect of a magnetic field has been explored in great detail~\cite{jiang2011possible,janssen2016honeycomb,yadav2016kitaev,gohlke2018dynamical,zhu2018robust,hickey2019emergence,nasu2018successive,liang2018intermediate,jiang2018field,yoshitake2019majorana,gordon2019theory,tanaka2020thermodynamic}, motivated by intriguing experiments on the leading Kitaev material candidate, $\alpha$-RuCl$_3$~\cite{kasahara2018majorana,yokoi2020half}. One finds that in a strong pre-dominantly in-plane magnetic field the thermal Hall response, $\kappa_{xy}/T$, appears to be \emph{half}-quantized, suggesting a propagating chiral edge mode with central charge $c=1/2$~\cite{readgreen2000}. This is consistent with expectations from the pure Kitaev model, where a small magnetic field produces a non-Abelian chiral spin liquid phase~\cite{kitaev2006}, with a single chiral Majorana edge mode (see Refs.~[\onlinecite{ye2018quant},\onlinecite{vinkler2018appox}] for some subtleties).

The effect of an \emph{electric} field on the Kitaev spin liquid is much less well-understood. Its response is encoded in the (effective) electric polarization operator~\cite{batista1} appropriate for Kitaev materials, as was worked out in detail in Refs.~[\onlinecite{miyahara2016theory,bolens1,bolens2}]. Using these polarization operators, the dielectric response can be computed~\cite{bolens1}, providing a natural route to subgap response in the optical conductivity. However, only the dynamic (linear) response has been considered, leaving open the question: how does the Kitaev spin liquid respond to a \emph{static} electric field?

\begin{figure}[t]
    \centering
    \includegraphics[width=\columnwidth]{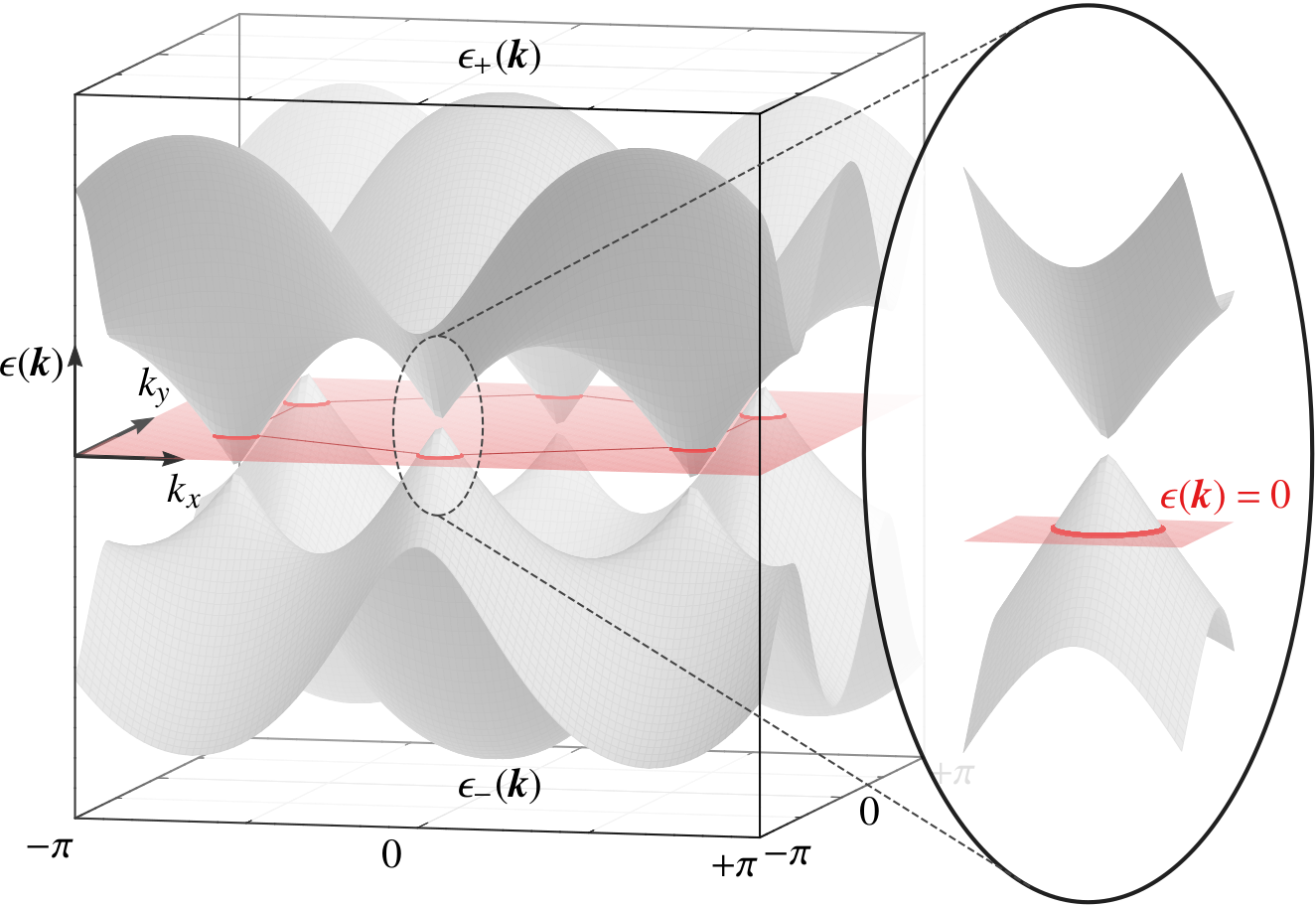}
    \caption{
    Illustration of the spectrum of the Kitaev model in the presence of both electric ($\vec{E}$) \emph{and} magnetic ($\vec{B}$) fields. An effective chemical potential is generated by the combination $\vhat{n} \cdot(\vec{E} \times \vec{B})$ where $\vhat{n}$ is the direction perpendicular to the honeycomb plane. Near the Dirac points this chemical potential induces Majorana-Fermi surfaces (see inset).   
    }
    \label{fig:crystal}
\end{figure}

In this article, we address this question, considering the effect of combined magnetic \emph{and} electric fields on Kitaev's honeycomb model. Using a generic, symmetry constrained polarization operator~\cite{miyahara2016theory,bolens2}, we apply degenerate perturbation theory to compute the effective Hamiltonian to second-order in both the magnetic and electric fields. We find that while the pure electric and magnetic contributions do not fundamentally alter the Kitaev spin liquid at this order, the leading \emph{Magnetoelectric} effect induces a chemical potential at the Dirac touching points and give rise to a \emph{Majorana-Fermi surface}~\footnote{Also known as \emph{Bogoliubov}-Fermi surfaces in the context of superconductors~\cite{bogo1,bogo2,bogo3}.}. This gapless spin liquid phase has no instabilities with respect to arbitrary perturbations and manifests in signatures in thermodynamic quantities. We further explore the interplay of this second-order cross-term with third-order contribution from the magnetic field, which stabilizes the gapped chiral spin liquid phase. We find that at low temperature it gives rise to a finite, but non-quantized~\cite{qin2011hall}, thermal Hall response, $\lim_{T \rightarrow 0} \kappa_{xy}/T$, in the gapless phase. 

Magnetoelectric effects~\cite{multi0,multi1} in ferromagnets have long attracted intense interest due to the potential for electrical control of magnetism (and vice versa) and for a variety of applications in spintronics. Applications in frustrated magnets (and anti-ferromagnets more broadly~\cite{multi2}) are not as well explored~\cite{savary2016quantum}. However, several results have established the potential utility of electrical probes, from subgap optical response in gapless spin liquids~\cite{potter1,bolens1,subgap2} to allowing electric control of fractionalized excitations in spin ice materials~\cite{khomskii2012electric,etienne1}. Here we offer Magnetoelectric route to generating a Majorana-Fermi surface. Since this is due to application of external fields, this has several advantages over more drastic perturbations; avoiding, for example, some of the complications of the quantum chemistry involved with doping or application of pressure.

The appearance of such Majorana-Fermi surfaces in Kitaev-like models has been discussed in several contexts, however each case carries some fundamental difficulty. These include instability towards nodal phases~\cite{hermanns2014oct,hermanns2015spin} (with nodal lines or points), sensitivity to symmetry-allowed perturbations~\cite{zhang2019vison} (fine-tuning) or realization through models with very unconventional lattices or interaction terms~\cite{yao2009algebraic,baskaran2009exact,tikhonov2010,chua2011,lai2011power} or with site-dependent (staggered) applied magnetic fields~\cite{takikawa2019}. In contrast, our result offers a natural avenue towards a Majorana-Fermi surface in a model that is directly related to realistic models of Kitaev materials. With this in mind, we discuss what kind of electric field strengths would be necessary to observe this physics in an idealized realization of Kitaev's honeycomb model; we find that, while large, the required electric fields are not far outside experimental reach. 

The article is structured as follows: in Sec.~\ref{sec:polariz} we define the Kitaev model and outline the derivation of the symmetry allowed polarization operator. In Sec.~\ref{sec:review} we review the exact solution of the pure Kitaev model and its symmetries to establish our notation and conventions. Sec.~\ref{sec:effective} describes our main results, covering the derivation of the effective Hamiltonian to second-order in the magnetic field alone (Sec.~\ref{sec:mag-two}), in the electric and magnetic fields (Sec.~\ref{sec:elec-mag}) and in the electric field alone (Sec.~\ref{sec:electric-second-order}). This effective Hamiltonian is then solved in Sec.~\ref{sec:solution} using the Majorana representation, and its spectrum, including the appearance of the Majorana-Fermi surface, is discussed in Sec.~\ref{sec:surface}. In Sec.\ref{sec:results} we explore the dependence on the direction of the electric and magnetic fields, focussing on the competition between the gap-opening term at third-order in the magnetic field and the Majorana-Fermi surface inducing second-order contribution. Finally, in Sec.~\ref{sec:discussion} we discuss relation to some recent works on perturbed Kitaev models, the effects of Majorana interactions, and provide some rough estimates of electric field strengths required to realize this physics, before concluding in Sec.~\ref{sec:conclusion} with an outlook and some perspective on future directions.

\section{Electric polarization in Kitaev materials}
\label{sec:polariz}
To begin, we define an effective $j=1/2$ pseudo-spin Kitaev model on a honeycomb lattice, as might appear in an ideal Kitaev material~\cite{kitaev2006}
\begin{equation}
    K \sum_{\avg{ij}_\gamma}  S_{i}^{\gamma}S_{j}^{\gamma},
\end{equation}
where $\avg{ij}_\gamma$ are the (labelled) nearest-neighbour bonds (see Fig.~\ref{fig:hexagon}). Generically, in the presence of both an electric and magnetic field we expect this Hamiltonian to acquire two new terms, taking the form
\begin{equation}
\label{eq:original-model}
    K \sum_{\avg{ij}_\gamma}  S_{i}^{\gamma}S_{j}^{\gamma}-\vec{B} \cdot \vec{M} -\vec{E} \cdot\vec{P},
\end{equation}
where $\vec{S}_i \equiv \vec{\s}_i/2$ are the pseudo-spins, $\vec{B}$ is the magnetic field, $\vec{M}~\equiv~g\mu_B \sum_i \vec{S}_i$ is the magnetization operator, $\vec{E}$ is the electric field, and $\vec{P}$ is the electric polarization operator.

To proceed, we need to express the electric polarization operator, $\vec{P}$, in terms of the pseudo-spins appropriate for the Mott insulating regime. This can be done perturbatively from the atomic limit, as discussed in Ref.~[\onlinecite{batista1}]. For Kitaev materials specifically, the structure of these polarization operators was worked out from similar microscopic considerations in Refs.~\cite{miyahara2016theory,bolens1,bolens2}, taking into account the details of the physics of 4$d$ or 5$d$ transition metal oxides. Rather than embark on such a microscopic approach, we will follow the discussion in Ref.~[\onlinecite{bolens2}], and parametrize the polarization operator using a generic form only constrained by lattice symmetries.

First, let us quickly review the derivation of the form of the polarization operators allowed by symmetry. Since the electric polarization operator is time-reversal even, inversion odd and translational invariant, we can immediately see that it must take the form
\begin{equation}
    P^\mu \equiv \sum_{\avg{ij}_\gamma} \vec{p}^\mu_\gamma \cdot (\vec{S}_i \times \vec{S}_j) + \cdots,
    \label{eq:polariz}
\end{equation}
where (by convention) we order each bond $\avg{ij}$ so that $i$ belongs to sublattice A and $j$ to sublattice $B$. We have truncated the expansion of this operator to nearest neighbours and only two-spin terms due to the structure of the perturbative expansion~\cite{batista1}, with other contributions appearing at higher order in $t/U$.

To go further, we need to know the details of the lattice symmetries. We use the conventional basis for the pseudo-spins defined such that, in the idealized limit, the octahedral cage of ligands are located along the $\pm{\vhat{x}},\pm{\vhat{y}}, \pm{\vhat{z}}$ directions. For simplicity, both the electric and magnetic fields are also defined with respect to this basis.
Trigonal distortion of the ligand cage generally lowers the local site symmetry of the transition metal ion to $D_{3d}$ (ignoring any small monoclinic distortions). The remaining symmetries of the crystal are the $C_{3}$ symmetry along the direction perpendicular to the honeycomb plane, and the $C_2$ symmetries along the nearest neighbour bonds. 

The three-fold symmetry links the three components of $\vec{P}$, so we can simply focus on $P^z$, recovering $P^x$ and $P^y$ by applying $C_3$ rotations. This leaves nine parameters in $\vec{p}_\gamma^z$. Under the action of the bond aligned $C_2$ symmetries one has that $P^z \rightarrow -P^z$ for the $z$ bond. To proceed, define the bond directions $\vhat{u}_\gamma$ as
\begin{align}
    \vhat{u}_x &\equiv \frac{\vhat{y}-\vhat{z}}{\sqrt{2}}, &
    \vhat{u}_y &\equiv \frac{\vhat{z}-\vhat{x}}{\sqrt{2}}, &
    \vhat{u}_z &\equiv \frac{\vhat{x}-\vhat{y}}{\sqrt{2}}.
\end{align}
 We further define an orthonormal frame for each bond $(\vhat{u}_\gamma,\vhat{v}_\gamma,\vhat{w}_\gamma)$ where $\vhat{w}_\gamma = \vhat{\gamma}$ and $\vhat{v}_\gamma \equiv \vhat{w}_\gamma \times \vhat{u}_\gamma$. Under its respective bond $C_2$ symmetry one has that $\vhat{u}_\gamma$ is invariant while $\vhat{v}_\gamma$ and $\vhat{w}_\gamma$ change sign. The bond symmetry then implies that
$\vhat{u}_z \cdot \vec{p}^z_z = 0$ and
\begin{align}
\vhat{u}_z \cdot \vec{p}^z_x &= -\vhat{u}_z \cdot \vec{p}^z_y, &
\vhat{v}_z \cdot \vec{p}^z_x &= +\vhat{v}_z \cdot \vec{p}^z_y, &
\vhat{w}_z \cdot \vec{p}^z_x &= +\vhat{w}_z \cdot \vec{p}^z_y. \nonumber
\end{align}
These four relations then leave us with five parameters, which we denote as $m_1,\dots,m_5$. After some rearrangement we can write the terms in the polarization operator [Eq.~(\ref{eq:polariz})] as~\cite{bolens2}
\begin{align}
\label{eq:polariz-params}
    \vec{p}^\mu_\gamma \equiv
    m_1 \hat{u}^\mu_\gamma {\vhat{u}}_\gamma +
    \hat{v}^\mu_\gamma \left(m_2 {\vhat{v}}_\gamma + m_4 {\vhat{w}}_\gamma\right)+
    \hat{w}^\mu_\gamma \left(m_3 {\vhat{w}}_\gamma+m_5 {\vhat{v}}_\gamma\right).
\end{align}
The large number of free parameters allowed in the polarization operator echoes the same freedom in the generic exchange Hamiltonian (four parameters) due to the relatively low bond symmetry~\cite{rau2014}.

Before moving on to the effects of the electric field on the physics of the Kitaev model, let us quickly note that one might consider including contributions to the magnetization operator at next to leading order (three-spin terms), or so-called orbital contributions~\cite{natori2019}. However, since these both appear at higher order than the two-spin terms that appear in the polarization operator, we will neglect them here.

\section{Review of the exact solution}
\newdimen\R \R=1.75cm    
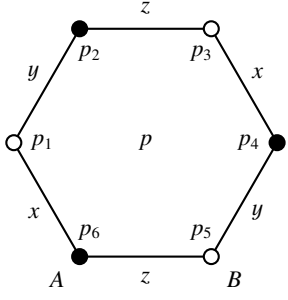
\begin{figure}[t]
    \centering
    \begin{tikzpicture}[baseline=-\R]
    \draw[thick] 
        (-0.5\R,-0.8660254\R) node[draw,circle,thick,inner sep=2pt,fill,label=below left:$A$]{} -- 
        (0.5\R,-0.8660254\R) node[draw,circle,thick,inner sep=2pt,fill=white,label=below right:$B$]{} -- 
        (\R,0)  node[draw,circle,thick,inner sep=2pt,fill]{} -- 
        (0.5\R,0.8660254\R) node[draw,circle,thick,inner sep=2pt,fill=white]{} --  
        (-0.5\R,0.8660254\R) node[draw,circle,thick,inner sep=2pt,fill]{} -- 
        (-\R,0) node[draw,circle,thick,inner sep=2pt,fill=white]{} -- 
        (-0.5\R,-0.8660254\R); 
    \node[label=center:$z$] at (0,1.03\R) {};
    \node[label=center:$y$] at (-0.85\R,0.53\R) {};
    \node[label=center:$x$] at (0.85\R,0.53\R) {};
    \node[label=center:$x$] at (-0.85\R,-0.53\R) {};
    \node[label=center:$y$] at (0.85\R,-0.53\R) {};   
    \node[label=center:$z$] at (0,-1.03\R) {};
    \node[label=center:$p$] at (0,0) {};
    \node[label=right:$p_1$] at (-\R,0) {};    
    \node[label=below:$p_2$] at (-0.425\R,0.8660254\R) {};
    \node[label=below:$p_3$] at (0.425\R,0.8660254\R) {};
    \node[label=left:$p_4$] at (\R,0) {};
    \node[label=above:$p_5$] at (0.425\R,-0.8660254\R) {};
    \node[label=above:$p_6$] at (-0.425\R,-0.8660254\R) {};
    \end{tikzpicture}
    \caption{Definition of the nearest-neighbour bond types, sublattices and the numbering convention for a hexagonal plaquette $p$.}
    \label{fig:hexagon}
\end{figure}

\label{sec:review}
We present here the structure of the exact solution which was originally proposed by Kitaev, as discussed in Ref.~\cite{kitaev2006}, to review the general ideas needed to discuss the effects of polarization and to establish our notation and conventions. 

Consider the  pure (isotropic) Kitaev model defined by the following Hamiltonian
\begin{equation}
    -J \sum_{\avg{ij}_\gamma}  \s_{i}^{\gamma}\s_{j}^{\gamma}.
\end{equation}
where we have transitioned to using the Pauli operators $\vec{\s}_i$, rather the spin-1/2 spins $\vec{S}_i$, with $J \equiv -K/4$ to simplify some of the later algebra. For simplicity we will assume $K<0$ so that $J>0$, without any loss of generality~\cite{kitaev2006}.
This model can be exactly solved, (partly) due to the large number of conserved ``flux'' operators that commute with both the Hamiltonian and with each other. We can define these flux operators for each honeycomb plaquette as a product of the spin operators going around the plaquette, with the spin component being that of the outward pointing bond type. Explicitly,
\begin{equation}
{W}_{p}=\s^{z}_{p_1}\s^{x}_{p_2}\s^{y}_{p_3}\s^{z}_{p_4}\s^{x}_{p_5}\s^{y}_{p_6},
\end{equation}
for a plaquette $p$ where the indexes start from the leftmost site and run clockwise (see Fig.~\ref{fig:hexagon}). It is easy to verify that when defined this way each flux operator commutes with every other flux operator and with the Hamiltonian, so that
\begin{align}
    [W_p,W_{p'}]&=0, & [H,W_p] &=0,
\end{align}
for any plaquettes $p$ and $p'$. Since ${W}_{p}^{2}=1$, each flux operator has eigenvalues given by $w_{p}=\pm 1$ and is a $\mathbb{Z}_2$ degree of freedom. Consequently, along with the Hamiltonian, they can be simultaneously diagonalized and the full Hilbert space can be decomposed into different flux sectors, each corresponding to the choice of the $W_p$ eigenvalues $\{{w}_{1},{w}_{2},...,{w}_{n}\}$. If there are $N$ sites, then there are $N/2$ plaquettes, and so fixing the flux sector halves the dimension of the Hilbert space, leaving $2^{N/2}$ degrees of freedom. The remaining ``half'' degree of freedom per site immediately suggests that these could be Majorana fermions.

To make this observation manifest, we follow \citet{kitaev2006} and write the Hamiltonian directly in terms of a Majorana representation
\begin{equation}
    \vec{\s}_i \equiv i \vec{b}_i c_i,
\end{equation}
where $\vec{b}_i \equiv (b^x_i,b^y_i,b^z_i)$ and $c_i$ are each Majorana fermions. To project back into the physical Hilbert space of the spins we must impose the constraint that $D_i \equiv b^x_i b^y_i b^z_i \f{c}_i = 1$. Practically, this allows for multiple representations of the spins, e.g $\vec{\sigma}_i \equiv D_i \vec{\sigma}_i = -i \vec{b}_i \times \vec{b}_i$. Since this constraint commutes with the Hamiltonian, the physical eigenstates of the spin model can be obtained from the eigenstates in the extended Hilbert space by projection. 

Using these Majorana fermions the Kitaev model in the extended Hilbert space can be written as
\begin{equation}
    i J \sum_{\avg{ij}_\gamma} \left(i b^\gamma_i b^\gamma_j\right) \f{c}_i \f{c}_j \equiv i J \sum_{\avg{ij}_\gamma} U_{ij} c_i c_j,
\end{equation}
where we have defined the $\mathbb{Z}_2$ gauge field operators $U_{ij} \equiv i b^\gamma_i b^\gamma_j$ on each link. Similar to the $\mathbb{Z}_2$ flux operators, these gauge field operators all commute with each other and with the Hamiltonian and thus can be simultaneously diagonalized in the extended Hilbert space. Since $U_{ij}^2 = 1$ their eigenvalues are simply $u_{ij} = \pm 1$. In terms of the $u_{ij}$ gauge-field the eigenvalues of the flux operators can be interpreted as the corresponding $\mathbb{Z}_2$ gauge flux
\begin{equation}
    {w}_{p}=u_{p_1 p_2}u_{p_2 p_3}u_{p_3 p_4}u_{p_4 p_5}u_{p_5 p_6}u_{p_6 p_1}.
\end{equation}
The extended Hilbert space then decomposes into spaces where the $u_{ij}$ are fixed and the effective Hamiltonian is a free Majorana problem given by
\begin{equation}
H[u] \equiv  J \sum_{\avg{ij}_\gamma} i u_{ij} c_i c_j .
\end{equation}
One can show~\cite{kitaev2006,lieb1994} that the ground state sector is such that the $\mathbb{Z}_2$ gauge fluxes are equal to one and thus the $u_{ij}$ can be chosen to be uniform with $u_{ij}=+1$ when going from the $A$ sublattice to the $B$ sublattice (up to gauge redundancy). This describes free Majoranas on a honeycomb lattice with only nearest-neighbour hopping; the spectrum is thus identical to that of graphene~\cite{grapheneRMP}, with the dispersion having linear touching points at the corners of the Brillouin zone.

This can be made precise by defining the Fourier transformed operators on each sublattice
\begin{align}
    c_{\vec{r},\alpha} = \frac{1}{\sqrt{N}} \sum_{\vec{k}} e^{i\vec{k} \cdot \vec{r}} c_{\vec{k},\alpha},
\end{align}
where we have defined each site $i$ by a unit cell $\vec{r}$ and a sublattice $\alpha= A$ or $B$ and we note that $\h{c}_{\vec{k},\alpha} = \f{c}_{-\vec{k},\alpha}$.
The free Majorana Hamiltonian for the ground state sector is then
\begin{equation}
\frac{1}{2}\sum_{\vec{k}>0} \left(c_{-\vec{k},A}\ c_{-\vec{k},B}\right)\left(
    \begin{array}{cc}
         0 & f(\vec{k}) \\
         \cc{f(\vec{k})} & 0
    \end{array}
    \right)\left(\begin{array}{c}
         c_{\vec{k},A}  \\
         c_{\vec{k},B} 
    \end{array}
    \right),
\end{equation}
where the sum runs over half the Brillouin zone and we have defined
\begin{equation}
\label{eq:defn-f}
    f(\vec{k}) \equiv 2iJ\left(1 + e^{-i\vec{k}\cdot\vec{a}_1}+e^{-i\vec{k}\cdot\vec{a}_2}\right),
\end{equation}
where $\vec{a}_1 \equiv (3\vhat{x} + \sqrt{3}\vhat{y})/2$, $\vec{a}_2 \equiv  (3\vhat{x} - \sqrt{3}\vhat{y})/2$ are the basis vectors of the honeycomb lattice. By diagonalizing this matrix we obtain the spectrum $\epsilon(\vec{k}) \equiv \pm |f(\vec{k})|$. As in graphene, we can expand $f(\vec{k})$ about $\vec{k} = \pm \vec{K}$, where $\vec{K} =  2\pi/3(\vhat{x}+\vhat{y}/\sqrt{3})$ is a corner of the Brillouin zone, to obtain
\begin{equation}
    f(\pm \vec{K} + \vec{q}) \approx -3J(q_x \pm i q_y) + O(q^2).
\end{equation}
One thus has a linear spectrum $\epsilon(\vec{K} + \vec{q}) \approx \pm v |\vec{q}|$, with Dirac velocity $v \equiv 3J$.

\subsection{Projective Symmetries}
\label{sec:projective}
Before moving on to the effects of perturbations on the Kitaev liquid,
we first review how symmetries of the spin model act in this Majorana basis.
We focus our attention on the constraints imposed by inversion symmetry (broken by an electric field) and time-reversal symmetry (broken by a magnetic field) within the ground state flux sector. The key property of both the symmetries is that they are implemented \emph{projectively}~\cite{kitaev2006,burnell2011,obrien2016}, that is the application of the symmetry operation must be followed by a $\mathbb{Z}_2$ gauge transformation. 

Consider first time-reversal: note that time-reversal, $\mathcal{T}$, is anti-unitary and maps $\vec{\s}_i \rightarrow -\vec{\s}_i$. In the Majorana representation a perfectly valid time-reversal operation is simply $\vec{b}_i \rightarrow \vec{b}_i$ and $c_i \rightarrow c_i$ with the imaginary prefactor giving the change in sign. However, this changes the link variables, as $u_{ij} \rightarrow -u_{ij}$ due to \emph{their} imaginary prefactor. Since we would like to work within the fixed gauge  sector with uniform $u_{ij}=+1$ we can undo this via a gauge transformation with a staggered sublattice sign. Our final (effective) time-reversal operator is then
\begin{equation}
c_i \xrightarrow{\mathcal{T}} (-1)^i c_i,
\end{equation}
where $(-1)^i$ is $+1$ on the $A$ sublattice and $-1$ on the $B$ sublattice. 
The treatment of inversion symmetry, $\mathcal{I}$, is essentially the same: instead of being anti-unitary, it interchanges the two sublattices, changing the sign of $u_{ij}$ in the same way. The effective action of inversion is then
\begin{equation}
c_i \xrightarrow{\mathcal{I}} (-1)^i c_{\mathcal{I}(i)},
\end{equation}
where $\mathcal{I}(i)$ is the site to which $i$ is mapped to under inversion.

These symmetry operations constrain the terms that can be generated by
the electric and magnetic fields. For example, since $\vec{B}$ is odd under
time-reversal any free Majorana terms like $i c_i c_j$ that are generated must respect that symmetry. This means, e.g. that any $O(B^2)$ terms must connect different sublattices, while any $O({B}^3)$ terms must only connect the same sublattice. Similarly, all of the pure electric field terms must connect different sublattices, as they are all time-reversal even.

Inversion symmetry is less restrictive; though it does not necessarily preserve the pair of sites in question (unlike time-reversal), one can still make some (more limited) statements. For example, since the first and third nearest-neighbour bonds are preserved by inversion they cannot be generated at odd order in $\vec{E}$. For second nearest neighbour bonds, inversion only relates the hopping on one bond to the distinct inverted bond. Finally, let us mention the cross term, at $O(EB)$ -- the main focus of our work -- which is odd under both time-reversal \emph{and} inversion and thus appears first in the second neighbour bonds, but is distinct from the usual $O(B^3)$ contribution.

\section{Perturbation theory in electric and magnetic fields}
\label{sec:effective}
We now review the usual perturbative approach to obtaining effective
Hamiltonians for the ground state flux sector when small perturbations are added. To set the stage, we will first recap known results~\cite{kitaev2006} for the effects of a magnetic field at $O(B^2)$ in Sec.~\ref{sec:mag-two}. We then proceed to derive the $O(EB)$ contributions in Sec.~\ref{sec:elec-mag} and the $O(E^2)$ contributions in Sec.~\ref{sec:electric-second-order}. Relevant aspects of the $O(B^3)$ contributions are reviewed~\cite{kitaev2006} in App.~\ref{app:cubic}.

\subsection{Magnetic Field}\label{sec:mag-two}
The effect of magnetic field is encapsulated in the piece of the Hamiltonian
\begin{equation}
   -\vec{B} \cdot \vec{M} \equiv  -\sum_{i}\vec{h}\cdot \vec{\s}_{i},
\end{equation}
where we define the (reduced) magnetic field $\vec{h} \equiv g \mu_B \vec{B}/2$. The form of perturbation theory to be used is motivated by the observation that the action of a single-spin operator always changes the flux sector. To see this consider the effect of the single spin $\s^z_i$ acting on the system -- one small piece of the magnetization $M^z$.  Using the commutation relations of the spins, we can see that when acting on state from the ground state flux sector
\begin{subequations}
\label{eq:flux-gen}
\begin{align}
    {W}_{x}(\s_{i}^{z} \ket{\Psi}) 
    &= -\s_{i}^{z} {W}_{x}\ket{\Psi}
    =-\s_{i}^{z} \ket{\Psi}, \\
{W}_{y}(\s_{i}^{z} \ket{\Psi}) 
    &= -\s_{i}^{z} {W}_{y}\ket{\Psi}
    =-\s_{i}^{z} \ket{\Psi}, \\
{W}_{z}(\s_{i}^{z} \ket{\Psi}) 
    &= +\s_{i}^{z} {W}_{z}\ket{\Psi}
    =+\s_{i}^{z} \ket{\Psi},     
\end{align}
\end{subequations}
where $W_\mu$ is the plaquette operator opposite to the $\mu$-bond connected to site $i$. We thus see that we have added fluxes on a pair of hexagons that are connected to the site where we acted the spin operator.  When considering the effect of this perturbation, the virtual states generated will thus not be within the ground state flux sector, but will necessarily mix in the two- or higher-flux sectors. Graphically, the flux configurations generated by the three spin components can be illustrated as
\R=0.6cm      
\begin{subequations}
\label{eq:flux-mag}
\begin{align}\centering
\s^x_i \ket{\Psi} &:
    \begin{tikzpicture}[baseline=-\R]
    \draw[fill=black!15!white] (0:\R) \foreach \x in {60,120,...,359} {-- (\x:\R)}-- cycle (90:\R);
    \draw[xshift=-1.5\R,fill=black!15!white,yshift=-0.8660254\R] (0:\R) 
    \foreach \x in {60,120,...,359} {-- (\x:\R)}-- cycle (90:\R);           
    \draw[yshift=-1.73205\R] (0:\R) 
    \foreach \x in {60,120,...,359} {-- (\x:\R)}-- cycle (90:\R);
    \node[color=black,circle,draw,circle,inner sep=1.25pt,fill=white] at (-0.5\R,-0.8660254\R) {};
    \node at (-0.2\R,-1.3\R) {$\s^x_i$};
    \end{tikzpicture},\\
\s^y_i \ket{\Psi} &:
    \begin{tikzpicture}[baseline=-\R]
    \draw (0:\R) \foreach \x in {60,120,...,359} {-- (\x:\R)}-- cycle (90:\R);
    \draw[xshift=-1.5\R,fill=black!15!white,yshift=-0.8660254\R] (0:\R) 
    \foreach \x in {60,120,...,359} {-- (\x:\R)}-- cycle (90:\R);           
    \draw[yshift=-1.73205\R,fill=black!15!white] (0:\R) 
    \foreach \x in {60,120,...,359} {-- (\x:\R)}-- cycle (90:\R);
    \node[color=black,circle,draw,circle,inner sep=1.25pt,fill=white] at (-0.5\R,-0.8660254\R) {};
    \node at (-0.2\R,-0.3\R) {$\s^y_i$};
    \end{tikzpicture}, \\
    \s^z_i \ket{\Psi} &:
    \begin{tikzpicture}[baseline=-\R]
    \draw[fill=black!15!white] (0:\R) \foreach \x in {60,120,...,359} {-- (\x:\R)}-- cycle (90:\R);
    \draw[xshift=-1.5\R,yshift=-0.8660254\R] (0:\R) 
    \foreach \x in {60,120,...,359} {-- (\x:\R)}-- cycle (90:\R);           
    \draw[yshift=-1.73205\R,fill=black!15!white] (0:\R) 
    \foreach \x in {60,120,...,359} {-- (\x:\R)}-- cycle (90:\R);
    \node[color=black,circle,draw,circle,inner sep=1.25pt,fill=white,
    label=left:${\s^z_i}$] at (-0.5\R,-0.8660254\R) {};
    \end{tikzpicture},
\end{align}    
\end{subequations}
where a filled hexagon indicates that $w_p = -1$ on that plaquette and an empty hexagon indicates $w_p=+1$. 
    
We thus consider a form of quasi-degenerate perturbation theory, where we derive an effective Hamiltonian within the ground state flux sector. Given the dimension of this
Hilbert space, it also naturally admits a description in terms of a single Majorana per site, i.e. the $c_i$ fermions. Formulating this perturbation theory strictly requires consideration of the full multi-particle spectra of the virtual flux states -- which is a much more challenging task. We will instead follow the approach of \citet{kitaev2006} and assume that most of the weight in the virtual processes comes from the single-particle excitations, an assumption that has been made plausible by more recent numerical studies~\cite{gohlke2018dynamical,knolle2016dynamics}. Practically, this means that we will replace any resolvents in our perturbation theory with a single energy scale -- the relevant flux gap.

To see how this is carried out, define the projection operator $P_0$ that projects into the ground state flux sector. In a magnetic field the effective Hamiltonian would then be (at second order)
\begin{equation}
\nonumber
    P_0 H_0 P_0 - \vec{h} \cdot \sum_i P_0 \vec{\s}_i P_0 - \sum_{\mu\nu}\sum_{ij} \frac{h_\mu h_\nu}{\Delta}
    P_0 \s^\mu_i (1-P_0) \s^{\nu}_j P_0 + \cdots,
\end{equation}
where $H_0$ is the Kitaev model and we have used that the resolvent $R$ is approximately given by $R = (1-P_0)/\Delta$, with $\Delta$ being the gap to creating two neighbouring flux pairs. From Ref.~[\onlinecite{kitaev2006}], one can estimate this to be $\Delta \approx 0.2672|J| = 0.067|K|$. 
Since $P_0\vec{M}P_0=0$ due to the change in flux, we see the effective Hamiltonian is then
\begin{equation}
    H_{\rm eff} = P_0 H_0 P_0  - \sum_{\mu\nu}\sum_{ij} \frac{h_\mu h_\nu}{\Delta}
    P_0 \s^\mu_i \s^{\nu}_j P_0 + \cdots.
\end{equation}
At this order, from the above considerations of flux generation [Eq.~(\ref{eq:flux-gen})], we can further see that
\begin{equation}
    P_0\s^\mu_i \s^\nu_j P_0 = \delta_{\mu\nu}\left[\delta_{ij} P_0+
     \delta_{\avg{ij}_\mu} P_0\s^\mu_i \s^\mu_j P_0\right].
\end{equation}
The first term describes adding two fluxes by applying a spin $\s^\mu_i$ at one site, and removing them using the same operator yielding an unimportant constant. The second term describes removing the added fluxes by applying the nearest-neighbour $\s^{\mu}_{i+\mu}$ and yields something non-trivial. We thus obtain
\begin{equation}
    H_{\rm eff} =  
    -\sum_{\avg{ij}_\gamma}
    \left(J+ \frac{2 h^2_\gamma}{\Delta}\right) 
    P_0 \s^\gamma_i \s^\gamma_j P_0  + {\rm const.}+O(h^3).
\end{equation}
The leading effects of the field are thus to renormalize the Kitaev couplings to render them (potentially) anisotropic, depending on the field direction. The presence of these $O(h^2)$ contributions also implies a finite magnetic susceptibility~\cite{willans2010disorder} at zero temperature. Note that the factor of two arises as the operators adding and removing the flux are different, and thus can be applied in two different orders.

\subsection{Electric and Magnetic Field}
\label{sec:elec-mag}
We now consider the effects of the electric polarization operator, following the same perturbative scheme that we used for the magnetic field (Sec.~\ref{sec:mag-two}). The leading, and most interesting, term will be the combination of the electric and magnetic field at $O(EB)$.  The first contribution from the electric field alone appears at $O(E^2)$ and is somewhat more complicated, without changing the essential physics of the leading term; we will cover it in detail in Sec.~\ref{sec:electric-second-order}. To make the bookkeeping simpler, we introduce the notation
\begin{equation}
    \vec{P}\cdot\vec{E} = \sum_{\avg{ij}_\gamma}\left(\sum_{\mu}  E_\mu \vec{p}_\gamma^\mu\right) \cdot (\vec{S}_i \times \vec{S}_j) \equiv
    \sum_{\avg{ij}_\gamma} \vec{\varepsilon}_\gamma \cdot (\vec{\s}_i \times \vec{\s}_j) ,
\end{equation}
where we have defined $\vec{\varepsilon}_{\gamma} \equiv \sum_{\mu} E_\mu \vec{p}^\mu_\gamma/4$. Explicitly, in terms of the parameters of the polarization operator [Eq.~(\ref{eq:polariz-params})], we have 
\begin{subequations}
\label{eq:defn-ep}
\begin{align}
    \varepsilon^{\alpha}_{\alpha} &= \frac{1}{4}\left[
    m_3 E_\alpha + \frac{m_4}{\sqrt{2}}  (E_\beta + E_\gamma)\right],\\
    \varepsilon^{\beta}_\alpha &=\frac{1}{4}\left[
    \frac{m_5}{\sqrt{2}} E_\alpha+
    \frac{1}{2}\left\{
    \left(m_1+m_2\right)E_\beta +
    \left(m_1-m_2\right)E_\gamma 
    \right\} \right],
\end{align}
\end{subequations}
where $\alpha,\beta,\gamma$ are a permutation of $x,y,z$. 

We first need to confirm that key property of the magnetic field perturbation that motivated our quasi-degenerate perturbation theory: that the action of the polarization operator changes the flux sector. To see how this works, consider the contribution of a single $z$-bond $\avg{ij}_z$ to $\vec{E}\cdot \vec{P}$, which takes the form
\begin{equation}
    \varepsilon^x_z \left(\s^y_i \s^z_j - \s^z_i \s^y_j\right)
   +\varepsilon^y_z \left(\s^z_i \s^x_j - \s^x_i \s^z_j\right)
   +\varepsilon^z_z \left(\s^x_i \s^y_j - \s^y_i \s^x_j\right).    \nonumber
\end{equation}
From the structure of this term we can see that while two spin operators are involved
in each (in contrast to the single spin for the magnetic field), they are always \emph{different} spin components due to the anti-symmetry imposed by inversion symmetry. This means that the fluxes generated by one spin operator are not removed by the other -- if one enumerates all the possible combinations, one can see that the two pairs of fluxes must only share at most a single plaquette, and therefore we are left with a pair of fluxes, just like in the magnetic field case. This then immediately implies the first order term is zero, with $P_0 \vec{P} P_0 = 0$.

The specific combinations of fluxes can be directly inferred from Eq.~(\ref{eq:flux-gen}), but are most clearly illustrated graphically. We delineate two types of flux configurations: those that give rise to a pair of nearest neighbour fluxes (type I) and those that give second-neighbour fluxes (type II). For the $z$-bond discussed above, four of the operators give type I flux configurations, as illustrated below
\R=0.6cm        
\begin{subequations}
\label{eq:type-one}
\begin{align}\centering
\s^y_i \s^z_j \ket{\Psi} &:
    \begin{tikzpicture}[baseline=-\R]
    \draw[fill=black!15!white] (0:\R) \foreach \x in {60,120,...,359} {-- (\x:\R)}-- cycle (90:\R);
    \draw[xshift=1.5\R,yshift=-0.8660254\R] (0:\R) 
    \foreach \x in {60,120,...,359} {-- (\x:\R)}-- cycle (90:\R);           
    \draw[xshift=-1.5\R,fill=black!15!white,yshift=-0.8660254\R] (0:\R) 
    \foreach \x in {60,120,...,359} {-- (\x:\R)}-- cycle (90:\R);           
    \draw[yshift=-1.73205\R] (0:\R) 
    \foreach \x in {60,120,...,359} {-- (\x:\R)}-- cycle (90:\R);
    \node[color=black,circle,draw,circle,inner sep=1.25pt,fill,
    label=left:${\s^y_i}$] at (-0.5\R,-0.8660254\R) {};
    \node[color=black,circle,draw,circle,inner sep=1.25pt,fill,
    label=right:${\s^z_j}$] at (0.5\R,-0.8660254\R) {};
    \draw[black,very thick] (-0.5\R,-0.8660254\R) -- (0.5\R,-0.8660254\R);
    \end{tikzpicture},\\
\s^z_i \s^y_j \ket{\Psi} &:
    \begin{tikzpicture}[baseline=-\R]
    \draw (0:\R) \foreach \x in {60,120,...,359} {-- (\x:\R)}-- cycle (90:\R);
    \draw[xshift=1.5\R,fill=black!15!white,yshift=-0.8660254\R] (0:\R) 
    \foreach \x in {60,120,...,359} {-- (\x:\R)}-- cycle (90:\R);           
    \draw[xshift=-1.5\R,yshift=-0.8660254\R] (0:\R) 
    \foreach \x in {60,120,...,359} {-- (\x:\R)}-- cycle (90:\R);           
    \draw[yshift=-1.73205\R,fill=black!15!white] (0:\R) 
    \foreach \x in {60,120,...,359} {-- (\x:\R)}-- cycle (90:\R);
    \node[color=black,circle,draw,circle,inner sep=1.25pt,fill,
    label=left:${\s^z_i}$] at (-0.5\R,-0.8660254\R) {};
    \node[color=black,circle,draw,circle,inner sep=1.25pt,fill,
    label=right:${\s^y_j}$] at (0.5\R,-0.8660254\R) {};
    \draw[black,very thick] (-0.5\R,-0.8660254\R) -- (0.5\R,-0.8660254\R);
    \end{tikzpicture},\\    
\s^x_i \s^z_j \ket{\Psi} &:
    \begin{tikzpicture}[baseline=-\R]
    \draw (0:\R) \foreach \x in {60,120,...,359} {-- (\x:\R)}-- cycle (90:\R);
    \draw[xshift=1.5\R,yshift=-0.8660254\R] (0:\R) 
    \foreach \x in {60,120,...,359} {-- (\x:\R)}-- cycle (90:\R);           
    \draw[xshift=-1.5\R,fill=black!15!white,yshift=-0.8660254\R] (0:\R) 
    \foreach \x in {60,120,...,359} {-- (\x:\R)}-- cycle (90:\R);           
    \draw[yshift=-1.73205\R,fill=black!15!white] (0:\R) 
    \foreach \x in {60,120,...,359} {-- (\x:\R)}-- cycle (90:\R);
    \node[color=black,circle,draw,circle,inner sep=1.25pt,fill,
    label=left:${\s^x_i}$] at (-0.5\R,-0.8660254\R) {};
    \node[color=black,circle,draw,circle,inner sep=1.25pt,fill,
    label=right:${\s^z_j}$] at (0.5\R,-0.8660254\R) {};
    \draw[black,very thick] (-0.5\R,-0.8660254\R) -- (0.5\R,-0.8660254\R);
    \end{tikzpicture},\\
\s^z_i \s^x_j \ket{\Psi} &:
    \begin{tikzpicture}[baseline=-\R]
    \draw[fill=black!15!white] (0:\R) \foreach \x in {60,120,...,359} {-- (\x:\R)}-- cycle (90:\R);
    \draw[xshift=1.5\R,fill=black!15!white,yshift=-0.8660254\R] (0:\R) 
    \foreach \x in {60,120,...,359} {-- (\x:\R)}-- cycle (90:\R);           
    \draw[xshift=-1.5\R,yshift=-0.8660254\R] (0:\R) 
    \foreach \x in {60,120,...,359} {-- (\x:\R)}-- cycle (90:\R);           
    \draw[yshift=-1.73205\R] (0:\R) 
    \foreach \x in {60,120,...,359} {-- (\x:\R)}-- cycle (90:\R);
    \node[color=black,circle,draw,circle,inner sep=1.25pt,fill,
    label=left:${\s^z_i}$] at (-0.5\R,-0.8660254\R) {};
    \node[color=black,circle,draw,circle,inner sep=1.25pt,fill,
    label=right:${\s^x_j}$] at (0.5\R,-0.8660254\R) {};
    \draw[black,very thick] (-0.5\R,-0.8660254\R) -- (0.5\R,-0.8660254\R);
    \end{tikzpicture} .
\end{align}    
\end{subequations}
Note that these type I configurations are
only generated by the $m_1$, $m_2$ and $m_5$ parts of the polarization operator.

The remaining two operators, coming from $\vhat{z}\cdot(\vec{\s}_i \times \vec{\s}_j)$, and generated by the $m_3$ and $m_4$ polarization operators,
give the type II flux configurations
\begin{subequations}
\label{eq:type-two}
\begin{align}
\centering
\s^x_i \s^y_i \ket{\Psi} &:
    \begin{tikzpicture}[baseline=-\R]
    \draw (0:\R) \foreach \x in {60,120,...,359} {-- (\x:\R)}-- cycle (90:\R);
    \draw[xshift=1.5\R,fill=black!15!white,yshift=-0.8660254\R] (0:\R) 
    \foreach \x in {60,120,...,359} {-- (\x:\R)}-- cycle (90:\R);           
    \draw[xshift=-1.5\R,fill=black!15!white,yshift=-0.8660254\R] (0:\R) 
    \foreach \x in {60,120,...,359} {-- (\x:\R)}-- cycle (90:\R);           
    \draw[yshift=-1.73205\R] (0:\R) 
    \foreach \x in {60,120,...,359} {-- (\x:\R)}-- cycle (90:\R);
    \node[color=black,circle,draw,circle,inner sep=1.25pt,fill,
    label=left:${\s^x_i}$] at (-0.5\R,-0.8660254\R) {};
    \node[color=black,circle,draw,circle,inner sep=1.25pt,fill,
    label=right:${\s^y_j}$] at (0.5\R,-0.8660254\R) {};
    \draw[black,very thick] (-0.5\R,-0.8660254\R) -- (0.5\R,-0.8660254\R);
    \end{tikzpicture},\\
\s^y_i \s^x_i \ket{\Psi} &:
    \begin{tikzpicture}[baseline=-\R]
    \draw (0:\R) \foreach \x in {60,120,...,359} {-- (\x:\R)}-- cycle (90:\R);
    \draw[xshift=1.5\R,fill=black!15!white,yshift=-0.8660254\R] (0:\R) 
    \foreach \x in {60,120,...,359} {-- (\x:\R)}-- cycle (90:\R);           
    \draw[xshift=-1.5\R,fill=black!15!white,yshift=-0.8660254\R] (0:\R) 
    \foreach \x in {60,120,...,359} {-- (\x:\R)}-- cycle (90:\R);           
    \draw[yshift=-1.73205\R] (0:\R) 
    \foreach \x in {60,120,...,359} {-- (\x:\R)}-- cycle (90:\R);
    \node[color=black,circle,draw,circle,inner sep=1.25pt,fill,
    label=left:${\s^y_i}$] at (-0.5\R,-0.8660254\R) {};
    \node[color=black,circle,draw,circle,inner sep=1.25pt,fill,
    label=right:${\s^x_j}$] at (0.5\R,-0.8660254\R) {};
    \draw[black,very thick] (-0.5\R,-0.8660254\R) -- (0.5\R,-0.8660254\R);
    \end{tikzpicture}.
\end{align}
\end{subequations}
Note that we have not explicitly written the signs or pre-factors in these expressions, we are simply illustrating the flux content of the generated states. The related patterns for the other types of bonds can be inferred using 
the three-fold symmetry; for a $\mu$-bond, then the operators corresponding
to $\vhat{\nu}\cdot(\vec{\s}_i \times \vec{\s}_j)$ where $\nu \neq \mu$ give the type I configurations, while the operators from $\vhat{\mu}\cdot(\vec{\s}_i \times \vec{\s}_j)$ give the type II configurations.

To make this more explicit, write the $O(EB)$ correction as
\begin{equation}
    -g \mu_B \sum_{\mu\nu}  B_\mu E_\nu  \left[ P_0 M_\mu \left(\frac{1-P_0}{\Delta}\right)  P_\nu P_0 
    + {\rm h.c.}\right],
\end{equation}
where again we have used that the resolvent reduces to $R = (1-P_0)/\Delta$ for the type I intermediate states of Eq.~(\ref{eq:type-one}). Since $P_0 \vec{M} P_0=0$ we can see that this reduces to 
\begin{equation}
    -\sum_{\mu\nu}\sum_{\avg{ij}_{\gamma}}\sum_k 
    \frac{h_\mu \varepsilon^\nu_{\gamma} }{\Delta}\left[ P_0\s^\mu_k\left(\vec{\s}_i \times \vec{\s}_j\right)^\nu  P_0 
    + {\rm h.c.}\right].
\end{equation}
We can use our knowledge of the intermediate (type I) flux states to simplify this further; we will work out one case explicitly, deriving the rest using symmetry.

\begin{figure}
    \R=0.8cm    
    \centering
    \begin{tikzpicture}
    \draw[fill=black!15!white] (0:\R) \foreach \x in {60,120,...,359} {-- (\x:\R)}-- cycle (90:\R);
    \draw[very thick,red] (-0.5\R,-0.8660254\R)--(\R,0);
    \draw[xshift=1.5\R,fill=black!15!white,yshift=-0.8660254\R] (0:\R) 
    \foreach \x in {60,120,...,359} {-- (\x:\R)}-- cycle (90:\R);           
    \draw[xshift=-1.5\R,yshift=-0.8660254\R] (0:\R) 
    \foreach \x in {60,120,...,359} {-- (\x:\R)}-- cycle (90:\R);           
    \draw[yshift=-1.73205\R] (0:\R) 
    \foreach \x in {60,120,...,359} {-- (\x:\R)}-- cycle (90:\R);
    \node[color=black,circle,draw,circle,inner sep=1.5pt,fill,
    label=left:${\s^z_i}$] at (-0.5\R,-0.8660254\R) {};
    \node[color=black,circle,draw,circle,inner sep=1.5pt,fill,
    label=right:${\s^x_j}$] at (0.5\R,-0.8660254\R) {};
    \node[color=black,circle,draw,thick,circle,fill=white,inner sep=1.5pt,
    label=above right:${\s^y_k}$] at (\R,0) {};    
    \draw[black,very thick] (-0.5\R,-0.8660254\R) -- (0.5\R,-0.8660254\R);
    \end{tikzpicture} \hspace{0.05cm}
    \begin{tikzpicture}
    \draw[fill=black!15!white] (0:\R) \foreach \x in {60,120,...,359} {-- (\x:\R)}-- cycle (90:\R);
    \draw[very thick,red] (-0.5\R,-0.8660254\R)--(\R,0);
    \draw[xshift=1.5\R,yshift=-0.8660254\R] (0:\R) 
    \foreach \x in {60,120,...,359} {-- (\x:\R)}-- cycle (90:\R);           
    \draw[xshift=-1.5\R,yshift=-0.8660254\R] (0:\R) 
    \foreach \x in {60,120,...,359} {-- (\x:\R)}-- cycle (90:\R);           
    \draw[yshift=-1.73205\R,fill=black!15!white] (0:\R) 
    \foreach \x in {60,120,...,359} {-- (\x:\R)}-- cycle (90:\R);
    \node[color=black,circle,draw,circle,inner sep=1.5pt,thick,fill=white,
    label=left:${\s^z_i}$] at (-0.5\R,-0.8660254\R) {};
    \node[color=black,circle,draw,circle,inner sep=1.5pt,fill,
    label=right:${\s^x_j}$] at (0.5\R,-0.8660254\R) {};
    \node[color=black,circle,draw,circle,fill,inner sep=1.5pt,
    label=above right:${\s^y_k}$] at (\R,0) {}; 
    \draw[black,very thick] (\R,0) -- (0.5\R,-0.8660254\R);
    \end{tikzpicture}     
    \caption{Illustration of the two contributions to the $O(EB)$ part of the effective Hamiltonian for the $x$-type second nearest-neighbour bond from $i$ to $k$ (via $j$) shown in red. The piece of the electric polarization operator is indicated by a thick bond with filled circles, while the piece from the magnetic field is indicated by an open circle. The hexagons that carry flux excitations in the corresponding virtual state are indicated.}
    \label{fig:illus-eb}
\end{figure}
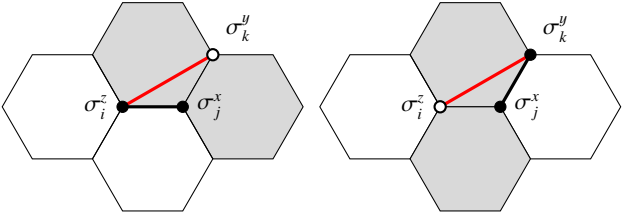

Focus on the contributions to a set of three sites ($i, j, k$) that define a second nearest neighbour $z$-bond type bond, as shown in Fig.~\ref{fig:illus-eb}. There are two processes that involve these three sites and both give a non-trivial contribution to the effective Hamiltonian. Together they give
\begin{align*}
   &= -\frac{2}{\Delta} P_0\left[ h_y \varepsilon^y_z \s^y_k \left(+\s^z_i\s^x_j\right) + h_z \varepsilon^z_y \s^z_i \left(-\s^y_k\s^x_j\right) \right]P_0, \\
   &= -\frac{2}{\Delta} \left(  h_y \varepsilon^y_z - h_z \varepsilon^z_y\right) P_0\s^z_i\s^x_j\s^y_k P_0,
\end{align*}
where the overall factor of two accounts for the Hermitian conjugate processes where the magnetic field is applied first.

This can be generalized and other bond types can be obtained by cyclically permuting the components of all vectors; one finds the final $O(EB)$ Hamiltonian to be
\begin{equation}
\label{eq:eb-spin}
    -\frac{2}{\Delta}\sum_{{{}^2\avg{ij}}_{\alpha(\beta)\gamma}} (-1)^i
   (h_\gamma \varepsilon^\gamma_{\alpha}-h_\alpha \varepsilon^\alpha_{\gamma})
     \epsilon_{\alpha\beta\gamma} P_0
    \s^\alpha_{i} \s^\beta_{i+\alpha}  \s^\gamma_{j} P_0 
\end{equation}
where ${}^2\avg{ij}_{\alpha(\beta)\gamma}$ indicates a second nearest-neighbour bond of type $\beta$ from $i$ to $j$, i.e. we get from $i$ to $j$ by traversing an $\alpha$-bond to an intermediate site, then on to $j$ via a $\gamma$-bond (see Fig.~\ref{fig:neighbours}). 

\subsection{Electric Field}
\label{sec:electric-second-order}
We now move to the processes that arise at second order due to the electric field alone. We identify three types of distinct processes: one that generates only two-spin interactions, renormalizing the nearest neighbour couplings and two further processes that generate four-spin interactions that link third and fourth nearest neighbours. These four-spin interactions are closely related to those introduced phenomenologically in Ref.~[\onlinecite{zhang2019vison}], see Sec.~\ref{sec:discussion} for a more detailed discussion.
\subsubsection{Two-Spin Process (Type II-Type II)}\label{sec:elec-two}
The first process of interest is generated by the type II flux configurations that did not contribute at $O(EB)$, as shown in Eq.~(\ref{eq:type-two}). Due to their geometrical arrangement, we see that only the operators that generate these further neighbour flux pairs [Eq.~(\ref{eq:type-two})] are the same ones that can remove them. For example, for a $\avg{ij}_z$ bond the two options are $\s^x_i \s^y_j$ and $\s^y_i \s^x_j$. Since using the identical operator simply results in a constant, only the cross terms give non-trivial contributions to the effective Hamiltonian; we write
\begin{align}
    &= \nonumber
    -\frac{(\varepsilon^z_z)^2}{\Delta'}P_0\left[
    \left(-\s^y_i \s^x_j\right)\left(+\s^x_i \s^y_j\right)+
    \left(+\s^x_i \s^y_j\right)\left(-\s^y_i \s^x_j\right)
    \right]
    P_0,\\
    &= +\frac{2(\varepsilon^z_z)^2}{\Delta'}P_0 \sigma^z_i \sigma^z_j P_0,
\end{align}
where $\Delta' \approx 0.2372|J| \approx 0.0593|K|$ is the gap for creating two further neighbour pairs~\cite{kitaev2006}. This can be done for each bond type, leading to the total effective Hamiltonian contribution
\begin{equation}
        +\sum_{\avg{ij}_\gamma}
        \frac{2(\varepsilon^\gamma_\gamma)^2}{\Delta'}
        P_0 \sigma^\gamma_i \sigma^\gamma_j P_0.
\end{equation}
Similar to the case of the $O(B^2)$ contributions, this simply renormalizes the bare Kitaev couplings, and (potentially) renders them anisotropic. 

Note that such processes do not exist for the type I flux configurations: the flux pair generated by each operator (on the same bond) is unique, and so the only way to remove them is by applying the original operator again. As in the type II case, this simply gives an unimportant constant.

\subsubsection{Four-Spin Process (Third Neighbour)}\label{sec:elec-four-a}
\begin{figure}[t]
    \R=0.8cm    
    \centering
    \begin{tikzpicture}
    \draw[fill=black!15!white] (0:\R) 
    \foreach \x in {60,120,...,359} {-- (\x:\R)}-- cycle (90:\R);
    \draw[xshift=1.5\R,yshift=-0.8660254\R] (0:\R) 
    \foreach \x in {60,120,...,359} {-- (\x:\R)}-- cycle (90:\R);           
    \draw[xshift=-1.5\R,yshift=-0.8660254\R] (0:\R) 
    \foreach \x in {60,120,...,359} {-- (\x:\R)}-- cycle (90:\R);           
    \draw[yshift=-1.73205\R,fill=black!15!white] (0:\R) 
    \foreach \x in {60,120,...,359} {-- (\x:\R)}-- cycle (90:\R);
    \draw[xshift=-1.5\R,yshift=-2.598076\R] (0:\R) 
    \foreach \x in {60,120,...,359} {-- (\x:\R)}-- cycle (90:\R);
    \draw[xshift=1.5\R,yshift=-2.598076\R] (0:\R) 
    \foreach \x in {60,120,...,359} {-- (\x:\R)}-- cycle (90:\R);    
    \draw[yshift=-3.4641\R] (0:\R) 
    \foreach \x in {60,120,...,359} {-- (\x:\R)}-- cycle (90:\R);  
    \draw[very thick,red] (-\R,-1.73205\R)--(\R,-1.73205\R);
    \node[color=black,circle,draw,circle,inner sep=1.5pt,fill,
    label=left:${\s^x_j}$] at (-0.5\R,-0.8660254\R) {};
    \node[color=black,circle,draw,circle,inner sep=1.5pt,fill,
    label=below left:${\s^y_i}$] at (-\R,-1.73205\R) {};
    \node[color=black,circle,draw,circle,inner sep=1.5pt,fill,
    label=right:${\s^y_k}$] at (0.5\R,-0.8660254\R) {};
    \node[color=black,circle,draw,circle,inner sep=1.5pt,fill,
    label=below right:${\s^x_l}$] at (\R,-1.73205\R) {};    
    \draw[black,very thick] (-0.5\R,-0.8660254\R) -- (-\R,-1.73205\R);
    \draw[black,very thick] (0.5\R,-0.8660254\R) -- (\R,-1.73205\R);
    \end{tikzpicture} \hspace{0.05cm}
    \begin{tikzpicture}
    \draw (0:\R) 
    \foreach \x in {60,120,...,359} {-- (\x:\R)}-- cycle (90:\R);
    \draw[xshift=1.5\R,yshift=-0.8660254\R] (0:\R) 
    \foreach \x in {60,120,...,359} {-- (\x:\R)}-- cycle (90:\R);           
    \draw[xshift=-1.5\R,yshift=-0.8660254\R] (0:\R) 
    \foreach \x in {60,120,...,359} {-- (\x:\R)}-- cycle (90:\R);           
    \draw[yshift=-1.73205\R,fill=black!15!white] (0:\R) 
    \foreach \x in {60,120,...,359} {-- (\x:\R)}-- cycle (90:\R);
    \draw[xshift=-1.5\R,yshift=-2.598076\R] (0:\R) 
    \foreach \x in {60,120,...,359} {-- (\x:\R)}-- cycle (90:\R);
    \draw[xshift=1.5\R,yshift=-2.598076\R] (0:\R) 
    \foreach \x in {60,120,...,359} {-- (\x:\R)}-- cycle (90:\R);    
    \draw[fill=black!15!white,yshift=-3.4641\R] (0:\R) 
    \foreach \x in {60,120,...,359} {-- (\x:\R)}-- cycle (90:\R);  
    \draw[very thick,red] (-\R,-1.73205\R)--(\R,-1.73205\R);
    \node[color=black,circle,draw,circle,inner sep=1.5pt,fill,
    label=left:${\s^y_r}$] at (-0.5\R,-2.598075\R) {};
    \node[color=black,circle,draw,circle,inner sep=1.5pt,fill,
    label=above left:${\s^x_i}$] at (-\R,-1.73205\R) {};
    \node[color=black,circle,draw,circle,inner sep=1.5pt,fill,
    label=right:${\s^x_s}$] at (0.5\R,-2.598075\R) {};
    \node[color=black,circle,draw,circle,inner sep=1.5pt,fill,
    label=above right:${\s^y_l}$] at (\R,-1.73205\R) {};    
    \draw[black,very thick] (-0.5\R,-2.598075\R) -- (-\R,-1.73205\R);
    \draw[black,very thick] (0.5\R,-2.598075\R) -- (\R,-1.73205\R);
    \end{tikzpicture}     
    \caption{Illustration of two contributions to the $O(E^2)$ part of the effective Hamiltonian for a four-spin coupling along a $z$-type third nearest-neighbour bond from $i$ to $l$ shown in red. The pieces of the electric polarization operator are indicated by thick bonds with filled circles. The hexagons that carry flux excitations in the corresponding virtual state are indicated.}
    \label{fig:illus-third}
\end{figure}
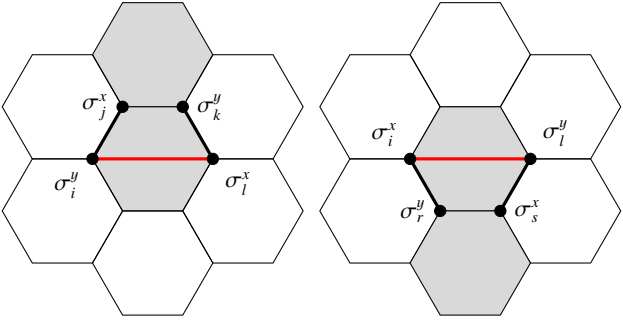

We now consider processes that involve the type I flux configurations, but with operators on different bonds. We first consider the process where the two pieces of the polarization operator are separated by a nearest neighbour bond and are non-parallel. Concretely, we can consider the processes illustrated in Fig.~\ref{fig:illus-third} that can be associated with a $z$-type third neighbour bond (see Fig.~\ref{fig:neighbours}). The first such process contributes (taking into account the reversed, or Hermitian conjugate, process as well)
\begin{equation}
\nonumber
    -\frac{2 \varepsilon^z_y \varepsilon^z_y}{\Delta}  P_0\left(+\s^x_j \s^y_i
    \right)\left(+\s^x_l \s^y_k\right)P_0
    = -\frac{2\varepsilon^z_y \varepsilon^z_y}{\Delta}   
    P_0 \s^y_i \s^x_j \s^y_k \s^x_l P_0,
\end{equation}
where $i, j, k, l$ are the four sites going clockwise along the top of the hexagon (see Fig.~\ref{fig:illus-third}). A similar process can be written for the sites running along the bottom of the hexagon, labeled $i, r, s, l$ going counter-clockwise, giving the final contribution
\begin{equation}
\nonumber
    = -\frac{2\varepsilon^z_x \varepsilon^z_y}{\Delta} \left(  
    P_0 \s^y_i \s^x_j \s^y_k \s^x_l P_0+
    P_0 \s^x_i \s^y_r \s^x_s \s^y_l P_0
    \right).
\end{equation}
Identical contributions exist for each third nearest-neighbour bond. We can then write
\begin{equation}
\nonumber
    -\frac{2}{\Delta}\sum_{{}^3\avg{ij}_{\alpha\beta(\gamma)}} 
    \varepsilon^\gamma_\alpha \varepsilon^\gamma_\beta
    \left(  
    P_0 \s^\beta_i \s^\alpha_{i+\beta} \s^\beta_{j+\alpha} \s^\alpha_j P_0+
    P_0 \s^\alpha_i \s^\beta_{i+\alpha} \s^\alpha_{j+\beta} \s^\beta_j P_0
    \right),
\end{equation}
where ${}^3\avg{ij}_{\alpha\beta(\gamma)}$ is a third-neighbour bond of type $\gamma$ (associated with the corresponding nearest-neighbour bond, see Fig.~\ref{fig:neighbours}). 
\subsubsection{Four-Spin Process (Fourth Neighbour)}
\label{sec:elec-four-b}
\begin{figure}[t]
    \R=0.8cm    
    \centering
    \begin{tikzpicture}
    \draw[fill=black!15!white] (0:\R) 
    \foreach \x in {60,120,...,359} {-- (\x:\R)}-- cycle (90:\R);
    \draw[xshift=1.5\R,fill=black!15!white,yshift=-0.8660254\R] (0:\R) 
    \foreach \x in {60,120,...,359} {-- (\x:\R)}-- cycle (90:\R);           
    \draw[xshift=-1.5\R,yshift=-0.8660254\R] (0:\R) 
    \foreach \x in {60,120,...,359} {-- (\x:\R)}-- cycle (90:\R);           
    \draw[yshift=-1.73205\R] (0:\R) 
    \foreach \x in {60,120,...,359} {-- (\x:\R)}-- cycle (90:\R);
    \draw[xshift=1.5\R,yshift=0.8660254\R] (0:\R) 
    \foreach \x in {60,120,...,359} {-- (\x:\R)}-- cycle (90:\R);
    \draw[xshift=3\R] (0:\R) 
    \foreach \x in {60,120,...,359} {-- (\x:\R)}-- cycle (90:\R); 
    \draw[very thick,red] (-0.5\R,-0.8660254\R)--(2\R,0);
    \node[color=black,circle,draw,circle,inner sep=1.5pt,fill,
    label=left:${\s^z_i}$] at (-0.5\R,-0.8660254\R) {};
    \node[color=black,circle,draw,circle,inner sep=1.5pt,fill,
    label=right:${\s^x_j}$] at (0.5\R,-0.8660254\R) {};
    \node[color=black,circle,draw,circle,inner sep=1.5pt,fill,
    label=left:${\s^x_k}$] at (\R,0) {};
    \node[color=black,circle,draw,circle,inner sep=1.5pt,fill,
    label=right:${\s^z_l}$] at (2\R,0) {};    
    \draw[black,very thick] (-0.5\R,-0.8660254\R) -- (0.5\R,-0.8660254\R);
    \draw[black,very thick] (\R,0) -- (2\R,0);
    \end{tikzpicture} 
    \caption{Illustration of a contribution to the $O(E^2)$ part of the effective Hamiltonian for a four-spin coupling along a $zy$-type fourth nearest-neighbour bond from $i$ to $l$ shown in red. The pieces of the electric polarization operator are indicated by thick bonds with filled circles. The hexagons that carry flux excitations in the corresponding virtual state are indicated.}
    \label{fig:illus-fourth}
\end{figure}
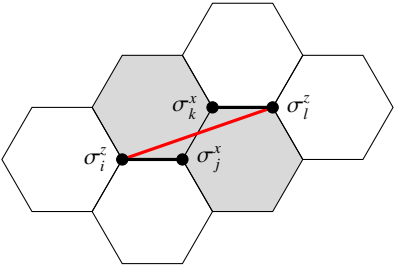

Finally, we consider the process where the two pieces of the polarization operator are separated by a nearest neighbour bond and are parallel. Explicitly, we can consider the processes illustrated in Fig.~\ref{fig:illus-fourth} that can be associated with a $zy$-type fourth neighbour bond (labeled using the composing nearest-neighbour bonds, see Fig.~\ref{fig:neighbours}). This contribution gives
\begin{equation}
\nonumber
    -\frac{2 (\varepsilon^y_z)^2}{\Delta}  P_0\left(+\s^z_i \s^x_j
    \right)\left(-\s^x_k \s^z_l\right)P_0
    = +\frac{2 (\varepsilon^y_z)^2}{\Delta}   
    P_0 \s^z_i \s^x_j \s^x_k \s^z_l P_0,
\end{equation}
where the path $i, j, k, l$ is shown in Fig.~\ref{fig:illus-fourth}. Unlike the previous type of process (Sec.~\ref{sec:elec-four-a}), this is the only contribution that involves these two endpoints. We thus can write the full set of contributions as
\begin{equation}
    +\frac{2}{\Delta}\sum_{{}^4\avg{il}_{\alpha\beta(\gamma)}}
    (\varepsilon^\beta_\alpha)^2
    P_0 \s^\alpha_i \s^\gamma_{i+\alpha} \s^\gamma_{j+\alpha} \s^\alpha_j P_0,
\end{equation}
where ${}^4\avg{ij}_{\alpha\beta(\gamma)}$ is an $\alpha\beta$-type fourth neighbour bond.

\section{Solution of effective Hamiltonian}
\label{sec:solution}
\R=0.65cm
\newdimen\mR \mR=-\R
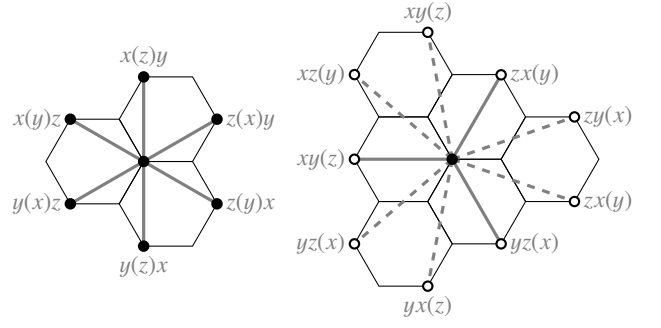
\begin{figure}
    \centering
    \begin{tikzpicture}[baseline=-4.2\R]
    \draw (0:\R) 
    \foreach \x in {60,120,...,359} {-- (\x:\R)}-- cycle (90:\R);
    \draw[xshift=-1.5\R,yshift=-0.8660254\R] (0:\R) 
    \foreach \x in {60,120,...,359} {-- (\x:\R)}-- cycle (90:\R);           
    \draw[yshift=-1.73205\R] (0:\R) 
    \foreach \x in {60,120,...,359} {-- (\x:\R)}-- cycle (90:\R);
    \draw[very thick,white!50!black] (-0.5\R,-0.8660254\R)--(\R,0) node [pos=1,right] {$z(x)y$}
    node[color=black,draw,circle,thick,inner sep=1.25pt,fill] {};
    \draw[very thick,white!50!black] (-0.5\R,-0.8660254\R)--(-0.5\R,0.8660254\R) node [pos=1,above] {$x(z)y$}
    node[color=black,draw,circle,thick,inner sep=1.25pt,fill] {};
    \draw[very thick,white!50!black] (-0.5\R,-0.8660254\R)--(-2\R,0) node [pos=1,left] {$x(y)z$}
    node[color=black,draw,circle,thick,inner sep=1.25pt,fill] {};
    \draw[very thick,white!50!black] (-0.5\R,-0.8660254\R)--(-2\R,-1.732\R) node [pos=1,left] {$y(x)z$}
    node[color=black,draw,circle,thick,inner sep=1.25pt,fill] {};
    \draw[very thick,white!50!black] (-0.5\R,-0.8660254\R)--(-0.5\R,-2.598\R) node [pos=1,below] {$y(z)x$}
    node[color=black,draw,circle,thick,inner sep=1.25pt,fill] {};
    \draw[very thick,white!50!black] (-0.5\R,-0.8660254\R)--(\R,-1.732\R) node [pos=1,right] {$z(y)x$} 
    node[color=black,draw,circle,thick,inner sep=1.25pt,fill] {};
    \node[color=black,circle,draw,circle,thick,inner sep=1.25pt,fill] at (-0.5\R,-0.8660254\R) {};
    \end{tikzpicture}   
    \begin{tikzpicture}
    \draw (0:\mR) 
    \foreach \x in {60,120,...,359} {-- (\x:\mR)}-- cycle (90:\mR);
    \draw[xshift=-1.5\mR,yshift=-0.8660254\mR] (0:\mR) 
    \foreach \x in {60,120,...,359} {-- (\x:\mR)}-- cycle (90:\mR);           
    \draw[yshift=-1.73205\mR] (0:\mR) 
    \foreach \x in {60,120,...,359} {-- (\x:\mR)}-- cycle (90:\mR);
    \draw[xshift=-1.5\mR,yshift=-2.598076\mR] (0:\mR) 
    \foreach \x in {60,120,...,359} {-- (\x:\mR)}-- cycle (90:\mR);
    \draw[xshift=-3\mR,yshift=-1.73205\mR] (0:\mR) 
    \foreach \x in {60,120,...,359} {-- (\x:\mR)}-- cycle (90:\mR);    
    \draw[yshift=-3.4641\mR] (0:\mR) 
    \foreach \x in {60,120,...,359} {-- (\x:\mR)}-- cycle (90:\mR);  
    \draw[very thick,white!50!black] (-\mR,-1.73205\mR)--(\mR,-1.73205\mR) node [pos=1,left] {$xy(z)$}
    node[color=black,draw,circle,thick,inner sep=1.25pt,fill=white] {};
    \draw[very thick,white!50!black] (-\mR,-1.73205\mR)--(-2\mR,0) node [pos=1,right] {$yz(x)$}
    node[color=black,solid,draw,circle,thick,inner sep=1.25pt,fill=white] {};
    \draw[very thick,white!50!black] (\R,1.73205\R)--(2\R,3.46\R) node [pos=1,right] {$zx(y)$}
    node[color=black,solid,draw,circle,thick,inner sep=1.25pt,fill=white] {};
    \draw[dashed,very thick,white!50!black] (\R,1.73205\R)--(-\R,0) node [pos=1,left] {$yz(x)$} 
    node[color=black,solid,draw,circle,thick,inner sep=1.25pt,fill=white] {};
    \draw[dashed,very thick,white!50!black] (\R,1.73205\R)--(-0.5\mR,0.8660254\mR) node [pos=1,below] {$yx(z)$}
    node[color=black,solid,draw,circle,thick,inner sep=1.25pt,fill=white] {};
    \draw[dashed,very thick,white!50!black] (\R,1.73205\R)--(\mR,-3.464\mR) node [pos=1,left] {$xz(y)$}
    node[color=black,solid,draw,circle,thick,inner sep=1.25pt,fill=white] {};
    \draw[dashed,very thick,white!50!black] (\R,1.73205\R)--(-0.5\mR,-4.330\mR) node [pos=1,above] {$xy(z)$}
    node[color=black,solid,draw,circle,thick,inner sep=1.25pt,fill=white] {};
    \draw[dashed,very thick,white!50!black] (\R,1.73205\R)--(-3.5\mR,-0.8660254\mR) node [pos=1,right] {$zx(y)$}
    node[color=black,solid,draw,circle,thick,inner sep=1.25pt,fill=white] {};
    \draw[dashed,very thick,white!50!black] (\R,1.73205\R)--(-3.5\mR,-2.598\mR) node [pos=1,right] {$zy(x)$}
    node[color=black,solid,draw,circle,thick,inner sep=1.25pt,fill=white] {};
    \node[color=black,circle,solid,draw,circle,thick,inner sep=1.25pt,fill] at (-\mR,-1.73205\mR) {};
    \end{tikzpicture}      
    \caption{Illustration of notation for the second (left), third and fourth nearest neighbour bonds (right) of the honeycomb
    lattice, as defined in Secs.~\ref{sec:elec-mag} and \ref{sec:electric-second-order}.}
    \label{fig:neighbours}
\end{figure}
With the effective Hamiltonian in the zero-flux sector worked out to second order in both the electric and magnetic fields, we now move on to the solution of this Hamiltonian using the Majorana representation.

The simplest terms are simply those that renormalize the nearest neighbour
couplings. It is useful to define the (induced) anisotropic Kitaev exchanges
$J_\gamma$ for each bond as
\begin{equation}
\label{eq:aniso-j}
    J_{\gamma} = J + 2\left(\frac{ h^2_\gamma}{\Delta} - \frac{(\varepsilon^\gamma_\gamma)^2}{\Delta'}\right),
\end{equation}
which includes contributions from the $O(B^2)$ and $O(E^2)$ processes (see Secs.~\ref{sec:mag-two} and \ref{sec:elec-two}). This modifies the function $f(\vec{k})$ that we encountered in the unperturbed solution [Eq.~(\ref{eq:defn-f})] to
\begin{equation}
    f_1(\vec{k}) \equiv 2i\left(J_z + J_y e^{-i\vec{k}\cdot\vec{a}_1} + J_x e^{-i\vec{k}\cdot\vec{a}_2}\right)= 2i \sum_\gamma J_{\gamma} e^{i\vec{k}\cdot(\vec{d}_\gamma-\vec{d}_z)}.
\end{equation}
with $\vec{d}_\alpha$ being the three (outward) nearest-neighbour bond directions of the honeycomb lattice, starting from the $A$ sublattice.

Next we consider the $O(EB)$ contributions: the three-spin term generated at this order [Eq.~(\ref{eq:eb-spin})] is a close analogue to the three-spin interaction that was obtained by \citet{kitaev2006} when going to \emph{third}-order in the magnetic field. However, a key difference is in the staggered sublattice sign which is required for the operator to be odd under inversion, and thus appear at $O(EB)$. These additional signs qualitatively change the effects of the interaction on the Majorana spectrum. Instead of opening a gap, as the $O(B^3)$ term does, this $O(EB)$ term will instead yield a one-dimensional manifold of zero energy states -- a \emph{Majorana} Fermi surface. 

To see this, start by writing the $O(EB)$ terms defined in Sec.~\ref{sec:elec-mag} [Eq.~(\ref{eq:eb-spin})] using the Majorana fermions as
\begin{align}
    P_0 
    \s^\alpha_{i} \s^\beta_{j}  \s^\gamma_{k} P_0 &=
    i^3 P_0 b^\alpha_i \f{c}_i   b^\beta_j  \f{c}_j  b^\gamma_k \f{c}_k P_0, \nonumber \\
    &=  -\epsilon_{\alpha\beta\gamma} 
    P_0  \left(i b^\alpha_i b^\alpha_j\right)\left(i b^\gamma_j  b^\gamma_k\right)
    \left(i \f{c}_i \f{c}_k\right) P_0 ,\nonumber \\
     &= -\epsilon_{\alpha\beta\gamma} 
    P_0  U_{ij} U_{jk} \left(i \f{c}_i \f{c}_k\right) P_0,
\end{align}
where we have made use of the constraint through the replacement
\begin{equation}
    b^\beta_j \f{c}_j  \equiv  \left(\epsilon_{\alpha\beta\gamma} b^\alpha_j  b^\beta_j b^\gamma_j \f{c}_j\right) b^\beta_j \f{c}_j  
    = \epsilon_{\alpha\beta\gamma} b^\alpha_j b^\gamma_j.
\end{equation}
 Now since $u_{ij}u_{jk} = -1$ in the zero-flux sector (due to bond orientations), the $O(EB)$ contribution to the effective Hamiltonian for the Majorana fermions is given by
\begin{equation}
    -\frac{2}{\Delta}\sum_{{{}^2\avg{ij}}_{\alpha(\beta)\gamma}} 
     (-1)^i (h_\gamma \varepsilon^\gamma_\alpha-h_\alpha \varepsilon^\alpha_\gamma)
       i \f{c}_i \f{c}_j.
\end{equation}
This is simply a second neighbour hopping, anisotropic in space and opposite in sign between the two sublattices. In Fourier space this can be written
\begin{equation}
    \frac{1}{2}\sum_{\vec{k}>0} g(\vec{k})  \left(
    c_{-\vec{k},A} c_{\vec{k},A}+
     c_{-\vec{k},B} c_{\vec{k},B}
    \right),
\end{equation}
where we have defined
\begin{align}
    g(\vec{k}) &=-\frac{8i}{\Delta} \sum_{\alpha \neq \gamma} 
        (h_\gamma \varepsilon^\gamma_\alpha-h_\alpha \varepsilon^\alpha_\gamma)
    e^{i \vec{k} \cdot(\vec{d}_\alpha -\vec{d}_\gamma)} , \nonumber\\
&=+\frac{16}{\Delta} \sum_{\alpha<\gamma} 
        (h_\gamma \varepsilon^\gamma_\alpha-h_\alpha \varepsilon^\alpha_\gamma)
         \sin\left[\vec{k}\cdot (\vec{d}_\alpha -\vec{d}_\gamma)\right].
\end{align}
We have used that, in general, the anti-symmetry of the Majorana operators imposes that $g(\vec{k}) = -g(-\vec{k})$, which is indeed satisfied by the above definition. Note that all of the other second order terms connect different sublattices, and so (at this order) this is the only contribution to $g(\vec{k})$. This is true \emph{generally} for the terms generated by the electric-field alone, given they are time-reversal even, they cannot provide any contributions to $g(\vec{k})$.

Finally, we consider the two types of four-spin interactions that are
generated at $O(E^2)$. Start first with the third-neighbour type (Sec.~\ref{sec:elec-four-a}), looking at the contributions to a $z$-type bond (as illustrated in Fig.~\ref{fig:illus-third})
\begin{align*}
    &= -\frac{2\varepsilon^z_x \varepsilon^z_y}{\Delta}
    P_0\s^y_i \s^x_j \s^y_k \s^x_l P_0, \\
    &=
    -\frac{2\varepsilon^z_x \varepsilon^z_y}{\Delta}P_0
    (i b^y_i c_i)(-ib^y_j b^z_j)(-i b^z_k b^x_k)(i b^x_l c_l)
    P_0 ,\\
    &=
    +\frac{2\varepsilon^z_x \varepsilon^z_y}{\Delta}u_{ij} u_{jk} u_{kl}
    P_0  (i c_i c_l) P_0 = +\frac{2\varepsilon^z_x \varepsilon^z_y}{\Delta}
    P_0  (i c_i c_l) P_0    .
\end{align*}
The same manipulations on the second process yield identical results. Generalizing to the full set of these four-spin terms, we therefore have the contribution
\begin{equation}
    -\frac{4}{\Delta} \sum_{{}^3\avg{ij}_{\alpha\beta(\gamma)}} 
    \varepsilon^\gamma_\alpha \varepsilon^\gamma_\beta (i c_i c_j),
\end{equation}
where we have reversed the sign, by ordering the bonds so that $i \in A$ and $j \in B$.
In Fourier space this gives a contribution to $f(\vec{k})$ of
\begin{equation}
\label{eq:f3}
   f_3(\vec{k}) \equiv -\frac{8 i}{\Delta}\sum_{\gamma}
    \varepsilon^\gamma_\alpha \varepsilon^\gamma_\beta e^{-i\vec{k} \cdot (2\vec{d}_\gamma + \vec{d}_z)},
\end{equation}
where the remaining indices are such that $\alpha, \beta  \neq \gamma$.
A similar procedure can followed for the fourth neighbour bonds (Sec.~\ref{sec:elec-four-b}); we simply quote the final result
\begin{equation}
     -\frac{2}{\Delta} \sum_{{}^4 \avg{ij}_{\alpha\beta(\gamma)}} 
        (\varepsilon^\beta_\alpha)^2 (i c_i c_j).
\end{equation}
Again, similarly, in Fourier space this gives a contribution to $f(\vec{k})$ that goes as
\begin{equation}
\label{eq:f4}
  f_4(\vec{k}) \equiv  -\frac{4i}{\Delta}\sum_{\alpha \neq \beta}
    (\varepsilon^\beta_\alpha)^2 e^{ i\vec{k} \cdot (2\vec{d}_\alpha - \vec{d}_\beta - \vec{d}_z)}.
\end{equation}
The final result for $f(\vec{k}) \equiv f_1(\vec{k}) + f_3(\vec{k}) + f_4(\vec{k})$ can be summarized as
\begin{align}
    \label{eq:full-f}
    f(\vec{k}) \equiv 2i\left[
    \sum_{\gamma} J_\gamma e^{i\vec{k}\cdot \vec{d}_\gamma }\right.
    &-\frac{4}{\Delta}\sum_{\gamma}
    \varepsilon^\gamma_\alpha \varepsilon^\gamma_\beta e^{-2 i\vec{k} \cdot \vec{d}_\gamma }  \nonumber \\
    &\left.
    -\frac{2}{\Delta}\sum_{\alpha \neq \beta}
    (\varepsilon^\beta_\alpha)^2 e^{ i\vec{k} \cdot (2\vec{d}_\alpha - \vec{d}_\beta)}
    \right]e^{-i\vec{k}\cdot\vec{d}_z},
\end{align}
where the $J_{\gamma}$ depend on the fields Eq.~(\ref{eq:aniso-j})
and the sums in the final two terms have the same meaning as in Eqs.~[\ref{eq:f3},\ref{eq:f4}].

The free Majorana Hamiltonian for the ground state sector, including the $O(B^2)$, $O(E^2)$ and $O(EB)$ contributions, is then given by
\begin{equation}
\frac{1}{2} \sum_{\vec{k}>0} \left(c_{-\vec{k},A}\ c_{-\vec{k},B}\right)\left(
    \begin{array}{cc}
         g(\vec{k}) & f(\vec{k}) \\
         \cc{f(\vec{k})} & g(\vec{k})
    \end{array}
    \right)\left(\begin{array}{c}
         c_{\vec{k},A}  \\
         c_{\vec{k},B} 
    \end{array}
    \right),
\end{equation}
The spectrum is then shifted from the case without the electric field
\begin{equation}
    \epsilon_{\pm}(\vec{k}) \equiv g(\vec{k}) \pm |f(\vec{k})|.
\end{equation}

It will be useful to include some aspects of the $O(B^3)$ contributions to the effective Hamiltonian into our analysis, so that we can explore the competition with the gap-opening terms. To this end we add the $O(B^3)$ second-neighbour hopping~\cite{kitaev2006}
\begin{equation}
\label{eq:three-spin-maj}
    -\frac{6 h_x h_y h_z}{\Delta^2}\sum_{{}^2\avg{ij}_{\alpha(\beta)\gamma}}
    \epsilon_{\alpha\beta\gamma}
    ic_i c_j.
\end{equation}
For details of the derivation of this term, see Appendix~\ref{app:cubic} or Ref.~[\onlinecite{kitaev2006}]. Note that we have not included the additional four-Majorana interaction term that is also generated at $O(B^3)$~\cite{kitaev2006}. In Fourier space this yields
\begin{equation}
\frac{1}{2} \sum_{\vec{k}>0} \left(c_{-\vec{k},A}\ c_{-\vec{k},B}\right)\left(
    \begin{array}{cc}
         g(\vec{k})+h(\vec{k}) & f(\vec{k}) \\
         \cc{f(\vec{k})} & g(\vec{k})-h(\vec{k})
    \end{array}
    \right)\left(\begin{array}{c}
         c_{\vec{k},A}  \\
         c_{\vec{k},B} 
    \end{array}
    \right), \nonumber
\end{equation}
where we have defined the new dispersion function
\begin{equation}
    h(\vec{k}) \equiv 
    +\frac{48 h_x h_y h_z}{\Delta^2}\sum_{\alpha(\beta)\gamma}
         \sin\left[\vec{k}\cdot (\vec{d}_\alpha -\vec{d}_\gamma)\right],
\end{equation}
where $\sum_{\alpha(\beta)\gamma}$ is defined as a sum over $\beta$ with $\epsilon_{\alpha\beta\gamma}=+1$. With this term included, the spectrum
of the Majorana fermions then takes the form
\begin{equation}
\label{eq:spectrum-full}
    \epsilon_{\pm}(\vec{k}) = g(\vec{k}) \pm \sqrt{h(\vec{k})^2 + |f(\vec{k})|^2}.
\end{equation}
With $\vec{E}=0$, and thus $g(\vec{k})=0$, this is the usual gapped spectrum expected for the Kitaev model in a small magnetic field~\cite{kitaev2006}. Expanding near the Dirac points, one has
\begin{equation}
    h(\vec{K}) = -h(-\vec{K}) = -\frac{72\sqrt{3} h_x h_y h_z}{\Delta^2}.
\end{equation}
The energy gap is then given by $2|h(\vec{K})|$ for $\vec{E}=0$; it will be useful to define a \emph{mass} of the Majorana fermions as $m \equiv |h(\vec{K})|$.

\section{Majorana Fermi surface}
\label{sec:surface}
How does the electric field perturbation affect the spectrum of the Majorana fermions, in particular the Dirac point? First, we should note that the $O(B^2)$ and $O(E^2)$ corrections preserve the Dirac touching, but shift them from $\pm \vec{K}$. Generically, this shift is also accompanied by the introduction of anisotropy and renormalization to the Dirac velocity as well. Consider the Dirac point near $\vec{K}$, schematically one has
\begin{equation}
    f(\vec{K}+\delta\vec{K}+\vec{q}) = -\left(v_1 [\mat{R}_{\theta} \vec{q}]_x + i v_2 [\mat{R}_{\theta} \vec{q}]_y\right) + O(q^2),
\end{equation}
where $\delta \vec{K}$ is the shift of the Dirac point, $\mat{R}_{\theta}$ is the rotation of the principal axes of the (anisotropic) Dirac cone, and $v_1$, $v_2$ are the two-independent Dirac velocities. The shift $\delta \vec{K}$ and the rotation angle $\theta$ are both second-order in the fields, while $v_n = v + O(E^2), O(B^2)$. In principle, the leading corrections can be worked out explicitly from an expansion of Eq.~(\ref{eq:full-f}), though we will leave this implicit for the sake of brevity.

For the $g(\vec{k})$ part things are somewhat simpler; since $\delta\vec{K} \sim O(B^2), O(E^2)$ and and $g(\vec{k})$ is already $O(EB)$, we only need to evaluate it at $\vec{K}$, with any corrections from $\delta \vec{K}$ being $O(EB^3)$ or $O(E^3 B)$. Evaluating this in terms of the physical electric and magnetic fields, $\vec{E}$ and $\vec{B}$, one thus has $g(\vec{K})$ being
\begin{align}
\label{eq:chem-pot}
g(\vec{K}) &=
    \frac{3 g\mu_B }{2\Delta}(m_1-m_2+\sqrt{2}m_5)
    \left[\vhat{n} \cdot (\vec{E} \times \vec{B}) \right]  ,
\end{align}
where $\vhat{n} = (\vhat{x}+\vhat{y}+\vhat{z})/\sqrt{3}$ is the direction perpendicular to the honeycomb plane. We thus see that the cross-term introduces a finite chemical potential at the Dirac points -- we define $\mu \equiv -g(\vec{K})$.  A similar argument applies for $h(\vec{k})$, since it is $O(B^3)$, any shifts due to the second-order terms only have higher order effects, and thus we can take $|h(\vec{K})| \sim 72\sqrt{3} h_x h_y h_z/\Delta^2$.

From the structure of $\vec{E}$ and $\vec{B}$ dependence we see that this chemical potential vanishes in several high symmetry configurations, including parallel electric and magnetic fields, as well as with either field being perpendicular to the honeycomb plane. This term is maximal when the electric and magnetic field are \emph{crossed}, i.e. $\vec{E}\cdot \vec{B} =0$, and both are in the honeycomb plane. We also see that the dependence field is distinct from other field induced terms, i.e. this term can remain finite when the gap-inducing $O(B^3)$ term vanishes. We also see that not all of the contributions to the polarization operator are effective in generating this chemical potential -- both $m_3$ and $m_4$ do not contribute as they only generate type II flux configurations.

The spectrum near the Dirac point is then given by
\begin{equation}
    \epsilon_{\pm}(\vec{K}+\delta\vec{K} + \vec{q}) \approx -\mu \pm 
    \sqrt{v^2|\vec{q}|^2+m^2},
\end{equation}
where we have ignored the renormalization of the Dirac velocity due to the $O(B^2)$ and $O(E^2)$ terms, setting $v = 3J$ for simplicity. Note that this mass does not necessarily render the system gapped; if small relative to the chemical potential it only opens a gap between the two Majorana bands. The low-energy excitations are now not at the Dirac touching point, but appear -- to a first approximation -- along a circular Majorana Fermi surface centered about the shifted Dirac points, with Fermi wave-vector
\begin{equation}
\label{eq:fermi-vec}
    q_F \equiv \frac{|\mu|}{v} = \frac{2|g(\vec{K})|}{3J} \sim O(EB).
\end{equation}
The Fermi velocity at this Majorana Fermi surface is inherited from the Dirac cone, with $v_F \sim v$, due to the linearity of the dispersion.  Note that we have ignored the $O(B^3)$ mass terms here, as they are parametrically smaller than the second-order terms that generate the chemical potential.

Just as for a more conventional Fermi surface, the additional low-energy excitations present qualitatively change the thermodynamic properties of this state at sufficiently low temperatures. For example, for temperatures well below the induced $O(EB)$ terms, the specific heat of the perturbed Kitaev model is $O(T)$, rather than the $O(T^2)$ in the original model, leading to a finite linear specific heat coefficient, $C/T \sim \gamma \neq 0$ as $T \rightarrow 0$.

In the next section, we will confirm this simple picture using a more complete calculation of the spectrum, as well as explore what happens when some of these terms vanish due to geometrical effects.

\section{Results}
\label{sec:results}

\begin{figure}[t]
    \centering
    \includegraphics[width=0.9\columnwidth]{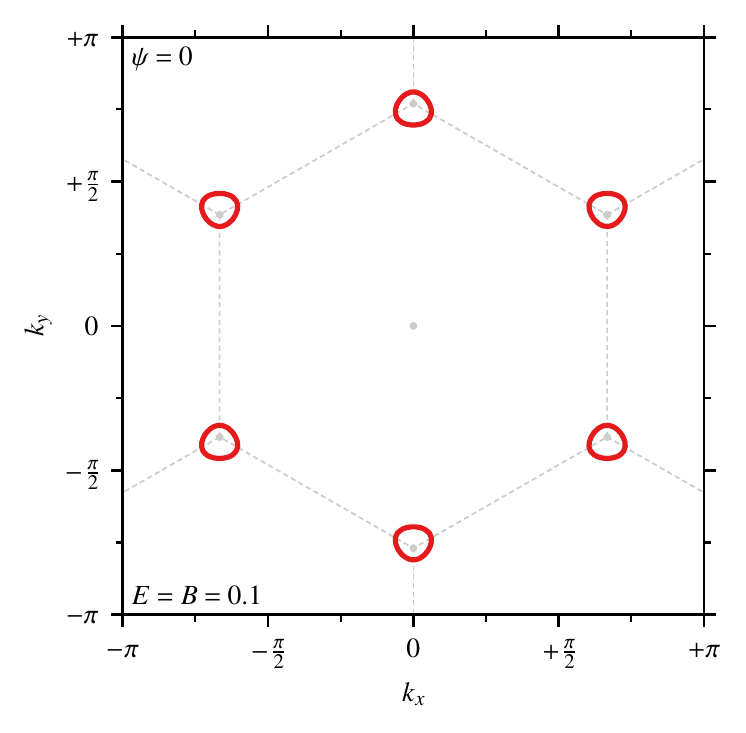}
    \hspace{0.05\columnwidth}
    \caption{Illustration of the Majorana-Fermi surface for crossed electric and magnetic fields, $\vec{E}=E\vhat{X}$ and $\vec{B} = B\vhat{Y}$ [Eq.~(\ref{eq:field-directions})], with $E=B=0.1$. See Sec.~\ref{sec:results} for the choice of electric polarization parameters and magnetic $g$-factors. }
    \label{fig:surf-only}
\end{figure}

For concreteness, we consider a configuration of the electric and magnetic fields that allows us to use the different geometrical dependences of the contributions to the effective Hamiltonian to isolate different aspects of the physics. We consider a configuration that can tune smoothly between regimes where the $O(B^3)$ contributions vanish, and those where the $O(EB)$ contributions vanish.  

For each case, we present the Majorana spectrum, and a few key physical observables, calculated (numerically) using the complete spectrum [Eq.~(\ref{eq:spectrum-full})].
For practical purposes we need to make some choices in our free parameters. First, we fix the polarization constants [Eq.~(\ref{eq:polariz-params})] to all be equal $m_1 =m_2 = m_3=m_4 = m_5 \equiv m_0$. This is a completely arbitrary choice, and is made simply to control the complexity of presentation -- we expect that the qualitative features of our results will not depend strongly on different, but still generic choices of the $m_n$. Next, we renormalize the energy scales and the electric and magnetic fields to correspond closely to the reduced variables (${h}$ and $\varepsilon$) that appear naturally in Sec.~\ref{sec:effective}. To this end, we set $J=1$, and take $g \mu_B/2 \equiv 1$,  absorbing the $g$-factor and Bohr magneton into the definition of $\vec{B}$. For the electric field, we take $m_0 \equiv 4$, absorbing their units into $\vec{E}$, and choosing the pre-factor to compensate for the corresponding factor in the definition of $\varepsilon$. 

\begin{figure}[t]
    \centering
    \includegraphics[width=0.8\columnwidth]{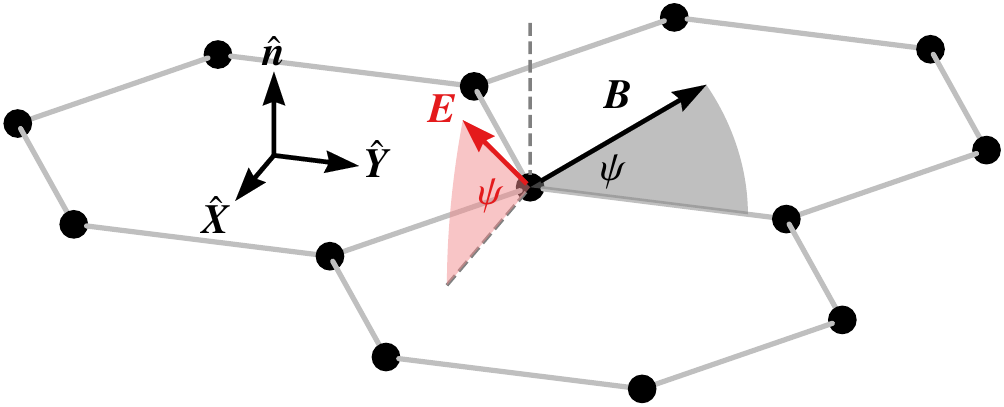}
    \caption{Illustration of the electric ($\vec{E}$) and magnetic field configurations ($\vec{B}$) considered in Sec.~\ref{sec:results}, relative to the honeycomb plane. These fields are parametrized by an angle $\psi$, as given in Eq.~(\ref{eq:field-directions}).
    }
    \label{fig:axes}
\end{figure}
More explicitly, the electric and magnetic fields are taken to be
\begin{subequations}
\label{eq:field-directions}
\begin{align}
    \vec{E} \equiv E\left(\cos{\psi} \vhat{X} + \sin{\psi} \vhat{n}\right),\\
    \vec{B} \equiv B\left(\cos{\psi} \vhat{Y} + \sin{\psi} \vhat{n}\right),
\end{align}
\end{subequations}
where $\vhat{X} = (2\vhat{z}-\vhat{x}-\vhat{y})/\sqrt{6}$ and $\vhat{Y}=(\vhat{x}-\vhat{y})/\sqrt{2}$. The orientation of these fields relative to the honeycomb plane is shown in Fig.~\ref{fig:axes}.
These choices yield a chemical potential
\begin{equation}
    \mu = \left(\frac{12\sqrt{2}}{\Delta} \right) EB \cos^2{\psi},
\end{equation}
which vanishes for $\psi=\pi/2$ and is maximal for $\psi=0$. 
Numerically, we expect then the radius of the Majorana-Fermi surface to
be $q_F \approx |\mu|/v \approx 31.76 \cdot EB \cos^2{\psi}$

The effective mass that splits the two Majorana bands then takes the form
\begin{equation}
        m = \left(\frac{24}{\Delta^2}\right) B^3 |3\sin{\psi} - 5\sin{(3\psi)}|,
\end{equation}
which vanishes for $\psi=0$ but is maximal for $\psi=\pi/2$. Note that this function also vanishes for $\psi_1 \approx 0.282047\pi$, in between these two limits.

\begin{figure}[t]
    \centering
    \includegraphics[width=\columnwidth]{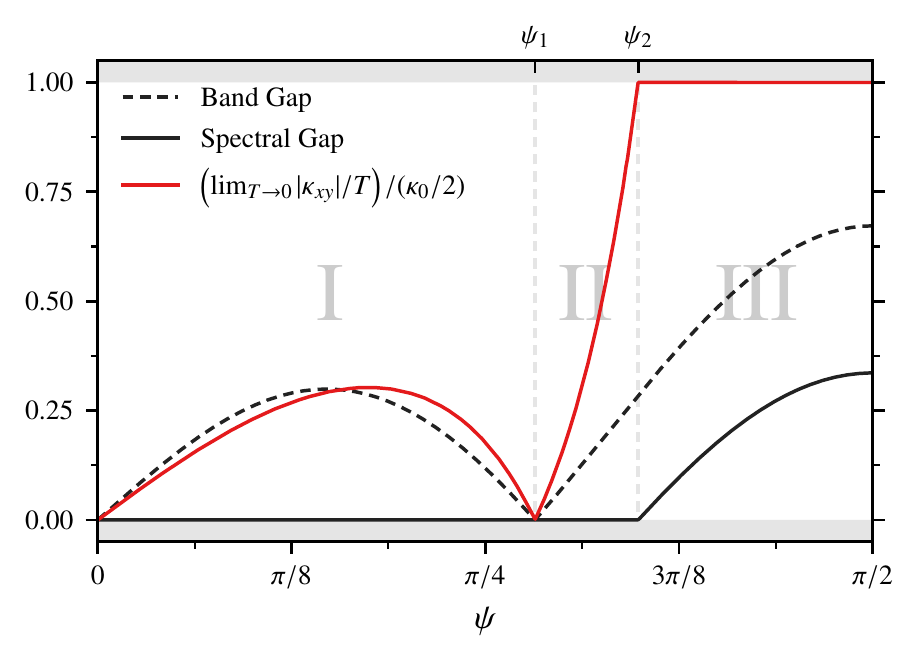}
    \caption{Illustration of the band gap ($\min_{\vec{k}}|\epsilon_+(\vec{k})-\epsilon_-(\vec{k})|$), spectral gap ($\min_{\vec{k},\pm} |\epsilon_{\pm}(\vec{k})|$) and thermal Hall conductivity [Eq.~(\ref{eq:hall})] as a function of the angle $\psi$ [Eq.~(\ref{eq:field-directions})] for $E=B=0.1$. 
    We identify three regions: region I where the band gap is finite and the spectral gap is zero, region II where the thermal Hall conductivity has changed sign and region III where the spectral gap has become finite. The boundary between regions I \& II is denoted as $\psi_1 \approx 0.282047\pi$, and the between II \& III is denoted as $\psi_2 \approx 0.3487\pi$.
    Sec.~\ref{sec:results} for the choice of electric polarization parameters and magnetic $g$-factors.
    }
    \label{fig:gap-plot}
\end{figure}

\begin{figure*}[t]
    \centering
    \includegraphics[width=\textwidth]{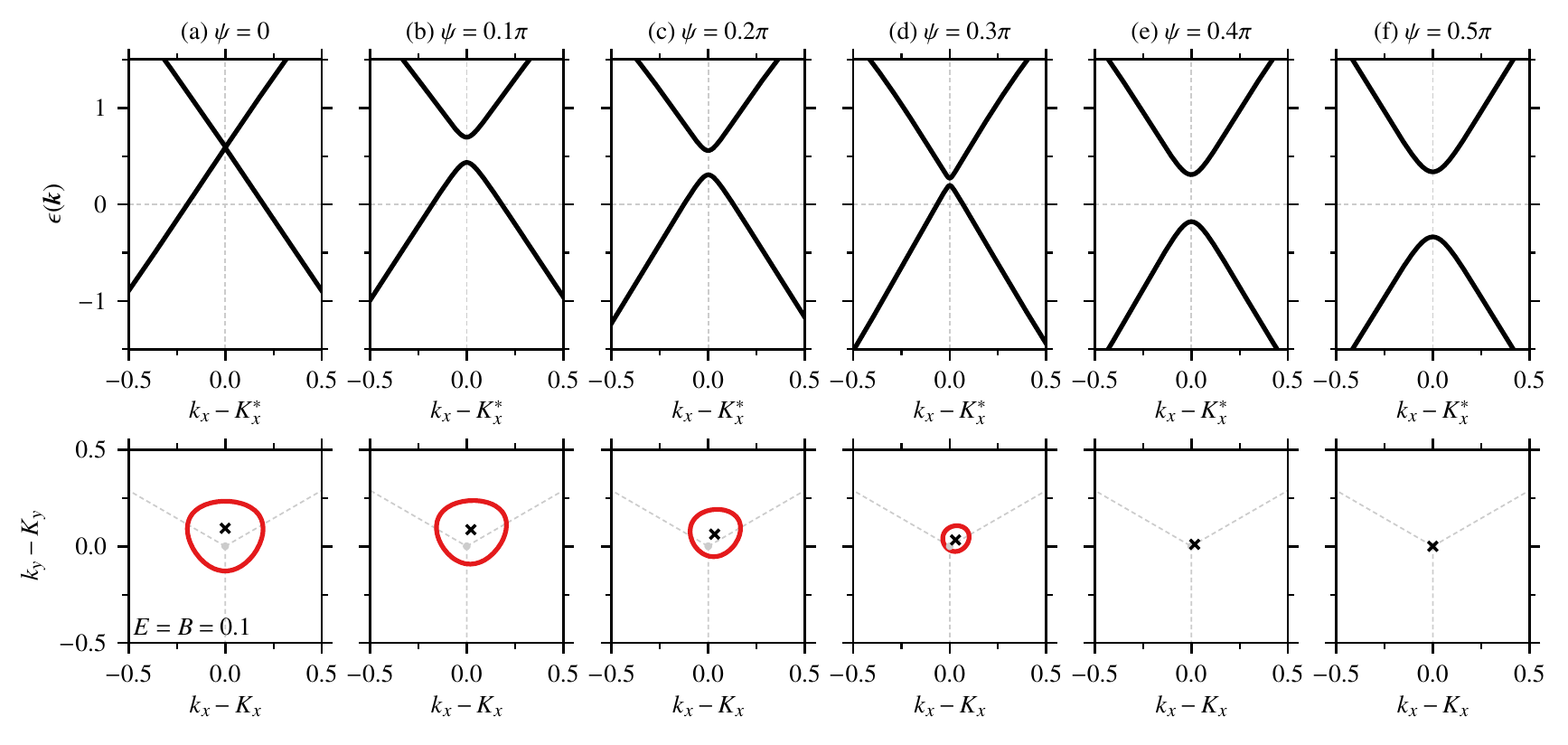}
    \caption{
    (a-f, top) Illustration of the spectrum of the Majorana fermions as a function of $\vec{E}$ and $\vec{B}$ (parametrized by $\psi$ [Eq.~(\ref{eq:field-directions})]) near the location of the (minimal) band gap at $\vec{K}^*$. 
    Gapless phases with Majorana-Fermi surfaces (a-d) and gapped phases (e,f) are shown.
    (a-f, bottom) Illustration of the Majorana-Fermi surfaces as a function of $\vec{E}$ and $\vec{B}$ (parametrized by $\psi$ [Eq.~(\ref{eq:field-directions})]). The location of the band gap ($\vec{K}^*$) is indicated. For $\psi > \psi_2 \approx 0.3487\pi$ (e,f) the system is gapped and there is no Majorana-Fermi surface.
    Throughout we choose field strengths $E=B=0.1$. See Sec.~\ref{sec:results} for the choice of electric polarization parameters and magnetic $g$-factors.
    }
    \label{fig:surf-cut}
\end{figure*}

We characterize the behaviour of the Majorana spectrum as a function of the angle $\psi$ by looking at two quantities: the spectral gap, which we define as the energy of the lowest lying excitation, (zero in the presence of a Majorana-Fermi Surface or Dirac point) and the band gap, which we define as minimum energy between the two Majorana bands (which is induced by the $O(B^3)$ from the magnetic field). Explicitly, we define the band gap as $\min_{\vec{k}}|\epsilon_+(\vec{k})-\epsilon_-(\vec{k})|$ and the spectral gap as $\min_{\vec{k},\pm} |\epsilon_{\pm}(\vec{k})|$. These quantities are shown in Fig.~\ref{fig:gap-plot} as a function of the angle $\psi$, with the Majorana-Fermi surface and spectrum near the band gap shown in Fig.~\ref{fig:surf-cut} for a handful of representative angles. Starting from $\psi=0$ where the chemical potential is maximal and the $O(B^3)$ term vanishes, both the spectral and band gaps are zero. For small $\psi$, the band gap becomes finite, while the spectral gap remains zero (labelled as region I). The band gap reaches a maximum as a function of $\psi$ before going to zero at the special value $\psi_1$ where the $O(B^3)$ term vanishes (the spectral gap is zero throughout). Increasing $\psi_1$ past this point, the Majorana-Fermi surface shrinks (region II) and then disappears near $\psi_2 \approx 0.3487\pi$, and the spectral gap then becomes finite (region III). A gapped spectrum, similar to what is found with only the magnetic field, is recovered as $\psi$ approaches $\pi/2$. 

These qualitative features can be diagnosed by looking at the behaviour of the thermal Hall conductivity of the Majorana-Fermions as $T\rightarrow 0$ (Fig.~\ref{fig:gap-plot}), computed via~\cite{qin2011hall} 
\begin{align}
\label{eq:hall}
    &\lim_{T \rightarrow 0}\frac{\kappa_{xy}}{T} = \frac{\pi^2}{3}\left(\frac{k_B^2}{\hbar}\right)\left\{
    \frac{1}{V}\sum_{\epsilon_n(\vec{k})<0}\ F^{(n)}_{xy}(\vec{k}) \right\}
    \equiv \kappa_0 \sum_n \Omega^{(n)},
\end{align}
where $F^{(n)}_{xy}=\partial_{x}A^{(n)}_{y}-\partial_{y}A^{(n)}_{x}$ is the Berry curvature of the $n^{\rm th}$ band, $V$ is the volume of the system and $A^{(n)}_{\mu}(k)=-i\langle n(k)|\partial_{\mu}|n(k)\rangle$ is the associated Berry potential. This can be 
naturally expressed in units of $\kappa_0 = \pi k_B^2/(6\hbar)$ and the total (occupied) Berry curvatures $\Omega^{(n)} \equiv 2\pi V^{-1}\sum_{\epsilon_n(\vec{k})<0}\ F^{(n)}_{xy}(\vec{k})$ for each band~\footnote{
For our numerical calculations, we compute the Berry curvature on finite lattices using the formulation of \citet{thermalhallfukui}, increasing the lattice size until convergence is reached for $\kappa_{xy}/T$.
}. When the spectrum is gapped, the $\Omega^{(n)}$ are the Chern numbers of each band and thus the thermal Hall response is quantized in units of $\kappa_0$.

In region I, with $\psi \neq 0$, the finite $O(B^3)$ contribution to the spectrum induces finite Berry curvature at the bottom of the two Majorana bands, near the location of the band gap. This Berry curvature gives a non-zero contribution to the thermal Hall conductivity, following Eq.~(\ref{eq:hall}). However, due the presence of the Majorana-Fermi surface the effectively ``occupied" states do not include the full Brillouin zone. Thus in region I $\kappa_{xy}/T$ is not integer or rational valued when expressed in units of $\kappa_0$, varying continuously as a function of $\psi$. At the special angle $\psi_1$ the $O(B^3)$ term vanishes, changing sign; the thermal Hall conductivity follows suit, remaining finite and unquantized in region II, but with opposite sign relative to region I. In region III, for $\psi > \psi_2$, there is no Majorana-Fermi surface, and the band gap is open and one thus recovers the half-quantized thermal Hall conductivity expected from Ref.~[\onlinecite{kitaev2006}].

\section{Discussion}
\label{sec:discussion}
We first discuss the relevance of these results to the body of related theoretical works on the Kitaev model and address the effects of interactions on the Majorana-Fermi surface (Sec.~\ref{sec:discussion:theo}). We then discuss some estimates for the electric fields that may be required to observe this physics, as well as some experimental hurdles in reaching these field strengths (Sec.~\ref{sec:discussion:exp}).
\subsection{Theoretical}
\label{sec:discussion:theo}
Let us begin with the relationship of the gapless states in this work to some recent studies in the literature. 
A set of four-spin interactions, similar to those derived in Sec.~\ref{sec:electric-second-order} at $O(E^2)$, were studied in Refs.~[\onlinecite{zhang2019vison},\onlinecite{zhang2020sixteen}].
These are not identical to those studied here; generically the four-spin interactions generated at $O(E^2)$ are anisotropic and there is a fixed relationship between the coefficients of the two different types of operators; however, given their similarity, some discussion is in order. 
In Ref.~[\onlinecite{zhang2019vison}] it is shown that for large couplings these operators lead to a change in the ground state flux sector of the model, stabilizing a rich variety of ``vison crystals''~\cite{zhang2019vison}. 
In our analysis, we have explicitly restricted ourselves to the zero-flux sector, assuming that perturbations from the electric and magnetic fields are sufficiently small to leave the flux sector unaffected.  The results of Ref.~[\onlinecite{zhang2019vison}] can give a rough idea of the range of validity of this approximation; using their notation, the coefficients of the four-spin terms should satisfy $K_3/K_1 \lesssim 0.1$ and $K_3'/K_1 \lesssim 0.4$.
In our notation, ignoring the (complicated) direction dependence of $\vec{E}$, this implies the loose criterion $ m_0^2 E^2/\Delta \lesssim 0.1$ (where $E$ is the electric field strength) to
preserve the zero-flux sector.
Alternatively, one may take the view that applying a strong electric field may be a route to stabilizing the ``vison crystal'' ground states described in Ref.~[\onlinecite{zhang2019vison}]. However, given this likely would push our perturbation theory to the limit of its regime of validity altogether, one must be cautious.

Some of these concerns could be addressed by more detailed numerical studies. For example, a numerical search for the true ground state flux sector~\cite{zhang2019vison} through quantum Monte Carlo studies at finite temperature~\cite{nasu2014vaporization,nasu2015thermal}. The degenerate perturbation theory itself could be validated using numerical techniques (such as DMRG or exact diagonalization) on the original model [Eq.~(\ref{eq:original-model})].

We also note the work of Ref.~[\onlinecite{takikawa2019}], which discusses the presence of Majorana-Fermi surfaces in a Kitaev model in the presence of site-dependent magnetic fields and off-diagonal exchanges $\Gamma$ and $\Gamma'$~\cite{rau2014}. In their proposal, these fields are generated by zigzag magnetic ordering in adjacent honeycomb layers, providing a source of both time-reversal and inversion symmetry breaking.
While magnetic ordering in some layers, with a Kitaev spin liquid in other layers may not seem particularly natural, the simultaneous breaking of these symmetries plays a similar role to the presence of $\vec{E}$ and $\vec{B}$ in this work.

\subsubsection{Stability}

An important question that must be addressed for the Majorana-Fermi surface we find in this work is stability to interactions. Similar states with gapless surfaces in three-dimensional Kitaev models~\cite{mfs2,mfs3,mfs4,mfs5,sim2020multipolar,bogomoon,link2020bogoliubov,jiang2020possible}, as ``Bogoliubov-Fermi" surfaces. For these systems, the absence of this kind of (effective) nesting symmetry precluded such an instability.

Generically, in the case of interest for the Kitaev model in electric and magnetic fields, all of the rotational symmetries of the system are broken, along with time-reversal symmetry. Thus, similar to the case of time-reversal breaking non-centrosymmetric superconductors, the system should be stable to the inclusion of weak short-range interactions. This can be seen from the symmetries of the problem: with finite, generic $\vec{E}$ and $\vec{B}$ the only remaining symmetries are discrete translations. Without any nesting vectors, interactions can only link a finite number of patches of the Majorana-Fermi surface and thus the enhancement that typically leads to instability is absent~\cite{shankarRMP,leeNFL}. Analogues of superconducting instabilities are also avoided~\footnote{This may be (alternatively) rationalized by noting that, while there are zero energy states at $\vec{k}$ and $-\vec{k}$ on the Majorana-Fermi surface, these are not independent and represent the usual Bogoliubov redundancy distributed in momentum space.}, given there is no $U(1)$ symmetry of the fermions left to spontaneously break~\cite{link2020bogoliubov}.

However, this argument has a few subtleties, given that the perturbations that generate the Majorana-Fermi surfaces also generate the interaction terms, just at higher order in perturbation theory. To see why this could complicate the analysis, note that in Sec.~\ref{sec:results} we showed that at second order in $E$ and $B$ the Majorana-Fermi surfaces are ellipses, which enjoy a kind of effective inversion symmetry through their centers. At higher-order, contributions would render these surfaces non-elliptical (see Fig.~\ref{fig:surf-only}), but would also include Majorana-Majorana interactions on equal footing. We leave the (potential) competition between these two effects, and thus the ultimate fate of these Majorana-Fermi surfaces to future studies.

We note that at the special angle $\psi_2$ where the spectral gap closes, the Majorana spectrum is that of a quadratic band touching, with a spectrum $\propto |\vec{k}|^2$. While the thermodynamic signature, $C \propto T$, this is the same as the Majorana-Surface, 
(due to its finite density of states) the effects of interactions may be different. A similar quadratic band touching point appears in AB stacked bilayer graphene~\cite{mccann2013electronic}. This has attracted attention due to its instability to weak short-range interactions, which leads to the emergence of a variety of new phases. We leave the discussion of the interaction effects on this quadratic Majorana spectrum, and any potential instabilities, to future work.

\subsubsection{Correlations}

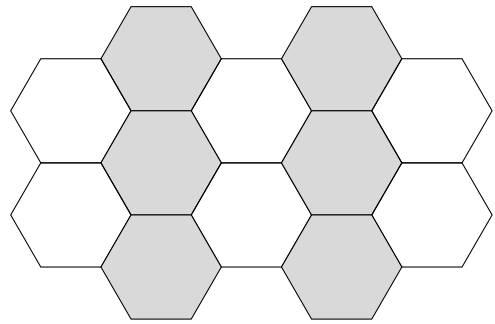
\begin{figure}
    \R=0.8cm
    \begin{tikzpicture}
    \draw[fill=black!15!white] (0:\R) 
    \foreach \x in {60,120,...,359} {-- (\x:\R)}-- cycle (90:\R);
    \draw[xshift=1.5\R,yshift=-0.8660254\R] (0:\R) 
    \foreach \x in {60,120,...,359} {-- (\x:\R)}-- cycle (90:\R);   
    \draw[xshift=-1.5\R,yshift=-0.8660254\R] (0:\R) 
    \foreach \x in {60,120,...,359} {-- (\x:\R)}-- cycle (90:\R);  
    \draw[fill=black!15!white,yshift=-1.73205\R] (0:\R) 
    \foreach \x in {60,120,...,359} {-- (\x:\R)}-- cycle (90:\R);
    \draw[xshift=-1.5\R,yshift=-2.598076\R] (0:\R) 
    \foreach \x in {60,120,...,359} {-- (\x:\R)}-- cycle (90:\R);
    \draw[xshift=1.5\R,yshift=-2.598076\R] (0:\R) 
    \foreach \x in {60,120,...,359} {-- (\x:\R)}-- cycle (90:\R);    
    \draw[fill=black!15!white,yshift=-3.4641\R] (0:\R) 
    \foreach \x in {60,120,...,359} {-- (\x:\R)}-- cycle (90:\R);  
    \draw[fill=black!15!white,xshift=3.0\R] (0:\R) 
    \foreach \x in {60,120,...,359} {-- (\x:\R)}-- cycle (90:\R);
    \draw[fill=black!15!white,xshift=3.0\R,yshift=-1.73205\R] (0:\R) 
    \foreach \x in {60,120,...,359} {-- (\x:\R)}-- cycle (90:\R);
    \draw[xshift=4.5\R,yshift=-0.8660254\R] (0:\R) 
    \foreach \x in {60,120,...,359} {-- (\x:\R)}-- cycle (90:\R); 
    \draw[xshift=4.5\R,yshift=-2.598076\R] (0:\R) 
    \foreach \x in {60,120,...,359} {-- (\x:\R)}-- cycle (90:\R);
    \draw[fill=black!15!white,xshift=3.0\R,yshift=-3.4641\R] (0:\R) 
    \foreach \x in {60,120,...,359} {-- (\x:\R)}-- cycle (90:\R);
    \end{tikzpicture}
    \caption{Illustration of $z$-type coverings. A product of $W_p$ over either the shaded plaquettes (or the unshaded plaquettes) yields the operator $\Sigma^{z}$ used in Eq.~(\ref{eq:mandal-condition}). The $\Sigma^x$ and $\Sigma^y$ operators are defined similarly, with the plaquettes in the product sharing $x$-type or $y$-type bonds.}
    \label{fig:flux-sigma}
\end{figure}

Finally, we note that the gapless nature of the spectrum in our model should lead to algebraic correlations in the spin-spin correlation functions. 
However, these may be ``hidden"~\cite{mandal2011,tish2011,song2016kitaev}; for example, as in the ideal Kitaev model the spin-spin correlations are ultra-short range in spite of the gapless Dirac spectrum of the Majorana fermions. 
Although the result of Ref.~[\onlinecite{mandal2011}] is not directly applicable in our effective Hamiltonian, it can be applied directly to the original model, Eq.~(\ref{eq:original-model}), including the electric polarization operator.  
Explicitly, using the criterion from Ref.~[\onlinecite{mandal2011}] if we consider the electric and magnetic field perturbation as $V \equiv -\vec{E} \cdot \vec{P} - \vec{B}\cdot\vec{M}$ from Eq.~(\ref{eq:original-model}), then
\begin{equation}
\label{eq:mandal-condition}   
[\Sigma^\mu,V] \neq 0,
\end{equation}
where $\Sigma^\mu$ is defined to be the product of flux variables over strings of plaquettes~\cite{mandal2011}, as shown in Fig.~\ref{fig:flux-sigma}. 
We thus expect that gaplessness of the Majorana-Fermi surface to manifest in power-law spin-spin correlations. 
Alternatively, this can be seen by computing the associated canonical transformation of the spin operators when carrying out the degenerate perturbation theory of Sec.~\ref{sec:effective}, as discussed in Ref.~[\onlinecite{song2016kitaev}] in a related context.


\subsection{Experimental}
\label{sec:discussion:exp}

We now address the potential applicability of these results to the growing family of ``Kitaev materials"~\cite{rau2016spin,winter2017models,hermanns2018physics} that are believed to have dominant Kitaev exchange. 
Ideally, one would like to apply electric and magnetic fields of magnitude that are realistically attainable and generate a Majorana-Fermi surface that could be observable in thermodynamic, transport or spectroscopic probes.

To do so, we must determine a reasonable range of estimates for the coefficients $m_i$ that appear in the (effective) electric polarization operator, as well as take a typical value for the Kitaev exchange $K$. 
The latter is straight forward: in transition metal Kitaev materials one typically expects $K \sim 5 \meV$~\cite{challengeswinter,winter2017models}. 
We thus expect $J = K/4 \sim 1.25\meV$ and thus the relevant flux gaps are $\Delta \sim \Delta' \sim 0.35 \meV$. 
Estimating the $m_i$ is more complex, with several distinct microscopic mechanisms potentially contributing at the same order~\cite{bolens1,bolens2}. 
In \citet{bolens2} these are estimated as the sum of two contributions: $m_n \equiv a \mathbb{A}_n + \mathbb{B}_n$ where $a$ is the nearest-neighbour distance (we will take it to be as in $\alpha$-RuCl$_3$, $a \sim 3.5\AA$~\cite{ruclcrystal}). 
The first contribution depends on the detailed choice of atomic and hopping parameters, but is estimated to be as large as $\mathbb{A} \sim 10^{-2} e a $~\cite{bolens2} where $e$ is the electron charge. 
The contribution $\mathbb{B}_n$ is not estimated in Ref.~[\onlinecite{bolens2}], so we will simply neglect it in this discussion. We thus have the estimate of $m_0 \sim 3.5 \cdot 10^{-2} e\AA = 3.5 \cdot 10^{-9} \meV/({\rm V}/{\rm m})$. 
This estimate is very rough and errors as large as an order of magnitude would not be surprising, given the uncertainty in the microscopic physics.

Given this value for $m_0$, we can now estimate the electric field strengths required to give a Majorana-Fermi surface of a given size. This size is set by the chemical potential which, for crossed fields (see Sec.~\ref{sec:results}) is $\mu \sim 3\sqrt{2} g m_0 \mu_B E B/(2\Delta)$ [restoring some constants in Eq.~(\ref{eq:chem-pot})]. Taking $g\sim 2$ and $B \sim 10\T$ we arrive at 
$$
\mu \sim \left\{2.4\cdot 10^{-8} \meV/({\rm V}/{\rm m})\right\} E.
$$
Since we are estimating $J \sim 1.25 \meV$ the Dirac velocity is $v \sim 3.75 \meV$ and so the radius of the Fermi surface is given by [Eq.~(\ref{eq:fermi-vec})]
$$
q_F a \sim \frac{|\mu|}{v} \approx 6.4 \cdot 10^{-9} \left(\frac{\rm V}{\rm m}\right)^{-1} E.
$$
To get a $q_F$ that is large, say $q_F a\sim 0.1$ we would thus need that $E \sim  10^7\ {\rm V}/{\rm m} \equiv E_0$. This is a very large field~\footnote{We also note that this is an optimistic estimate; a significantly larger $E_0$, for example of size $\sim 10^8-10^9\ {\rm V}/{\rm m}$ almost surely possible, depending on the precise estimates of the microscopic parameters $\mathbb{A}$, $\mathbb{B}$, the magnetic field strength $B$, and the Kitaev exchange $K$.}; though we note that static electric fields of of magnitude $\sim 10^6 - 10^7\ {\rm V}/{\rm m}$ have been used in studies of magnetic systems~\cite{largefield0,largefield1,largefield2,largefield3,largefield5}. 

This na\"ive analysis must be supplemented with a number of caveats. First, it must be self-consistent: we have that at these large fields $m_0 E_0~\sim~0.1 \meV$ which is not significantly smaller than the flux gap $\Delta~\sim~0.35\meV$, so our perturbation theory may be approaching the edges of its validity. Further, we have the requirement that we must remain in the zero-flux sector~\cite{zhang2019vison}. For this large electric field $m_0^2 E_0^2/(J\Delta) \sim 0.02$; reasonably far from where one might expect a transition to a new flux sector (see discussion in Sec.~\ref{sec:discussion}).~\footnote{We note that the size of the zero-flux sector may be quite different when we account for the anisotropy of our four-spin interactions (larger or smaller).} Similar considerations must be applied to the \emph{magnetic} field as well, concerning the validity of its perturbation theory and its effect on the phase boundary of the zero-flux phase. One route to lowering the required electric field would be via larger magnetic fields along a direction where the $O(B^3)$ terms vanish. Alternatively, one could search for Kitaev materials where $K$ or $\Delta$ are smaller; rare-earth Kitaev materials~\cite{rekitaev1,rekitaev2} may be promising alternatives, if the size of electric field coupling is not dramatically changed. Kitaev materials based on metal-organic frameworks~\cite{mof1}, which have longer bond-lengths, may also change the balance of the exchange and electric field interactions.

More practically, at large static fields such as these one must also be aware of effects on the physical system that are not included with our minimal model [Eq.~(\ref{eq:original-model})]. This includes potential structural distortions of the lattice along the field, the introduction of charge carriers through electrostatic doping as well as the modification of the microscopic exchange processes due to the field. More dramatically, the material itself could experience dielectric breakdown; for Mott insulators breakdown electric fields of order $\sim 10^8-10^9\ {\rm V}/{\rm m}$ are not atypical, given the expected relationship to the on-site repulsion~\cite{breakdown1}. One potential route to avoid some of these issues is to consider large electric field pulses, perhaps through the application of laser light, that can reach these field strengths over short time windows. Realizing this physics only within a short time window (long with respect to the spin dynamics) would however limit the experimental probes available to characterize the system.

An alternative route to large static electric fields could be through the engineering of heterostructures of Kitaev materials. Recently, heterostructures of graphene (single and bilayer) and (bulk and monolayer) $\alpha$-RuCl$_3$ have been reported in the literature~\cite{hetero1,hetero2,hetero3,hetero4}. In each case, the associated potential difference (large effective electric field) results in significant doping of both the graphene and the $\alpha$-RuCl$_3$. While the effect of these fields appears to be too strong for our purposes, the possibility of van der Waals heterostructures of Kitaev materials as a platform to explore the electric field physics discussed in this work remains an intriguing possibility. We note that several recent theoretical works have explored the (related) effect of tunnelling in such heterostructures~\cite{carrega2020tunnelling,pereira2020electrical,konig2020tunneling,feldmeier2020local} through the effect of \emph{localized} electric fields.

While each of these complications to achieving the large electric fields needed here are important, these need not be insurmountable. From the properties of the Majorana Hamiltonian these surfaces can appear whenever we break time-reversal and inversion symmetry simultaneously; combined electric and magnetic fields are simply the minimal perturbation that does so. Thus, while we have derived our results within a well-defined framework for the pure Kitaev model with a specific (though generic) electric polarization operator, we expect that generation of a Majorana-Fermi surface not to be strictly dependent on these implementation details. For example, physics similar to that introduced here may be introduced by disorder that breaks inversion symmetry. Explicitly, something akin to charged defects could introduce in-plane electric fields and, in regions where these fields are sufficiently uniform, our results would apply. Introducing a magnetic field would then generate a Majorana chemical potential in concert with these defect fields. We leave the exploration of these possibilities for future work.

Finally, let us make a connection to more conventional Magnetoelectric effects that have been extensively discussed in multiferroic materials. A first step to gauge the importance of these kinds of magnetoelectric interactions, before embarking on a detailed search for a Majorana-Fermi surface, will be to look at the bulk magnetization as a function of electric field strength. As in conventional Magnetoelectric materials, due to the presence of the $O(EB)$ terms in the effective Hamiltonian, one expects that at leading order (fixing directions and coupling coefficients)
$$
M \sim \chi B + \alpha E^{2}B,
$$
where $\chi$ is the susceptibility and $\alpha$ a Magnetoelectric coupling. A similar term, with $E$ and $B$ switched, would appear in the electric polarization. This effect could be measurable, even if making a sufficiently large enough Majorana-Fermi surface is not feasible, and could provide some guidance on the size of the unknown couplings, $m_n$, that appear with the polarization.

\section{Conclusion}
\label{sec:conclusion}

In summary, we have analyzed the effects of combined static electric and magnetic fields on Kitaev's honeycomb model. 
Starting from the symmetry-allowed effective polarization operator~\cite{bolens1,bolens2}, we used degenerate perturbation theory to derive an effective Hamiltonian to second-order in both the magnetic and electric fields. 
This effective Hamiltonian is solvable and describes a set of (free) Majorana fermions with a Majorana-Fermi surface over a wide range of parameters, including the neighbourhood of the Kitaev spin liquid. 
We explored the effects of the $O(B^3)$ gap-opening perturbation, showing how it competes with the generation of the Majorana-Fermi surface depending on the relative orientations of $\vhat{E}$ and $\vhat{B}$. 
The spectrum of this spin liquid with Majorana-Fermi surface was characterized in the weak field limit, where we derived the low-energy form of the dispersion, and determined the thermodynamic properties and thermal Hall coefficient.
Finally, we discussed some related work, speculated on the effects of interactions and provided rough estimates of the magnitude of electric and magnetic fields that are required to render the size of the Majorana-Fermi surface significant.

We hope that our results further motivate the use of electric fields as potentially useful perturbations in Kitaev materials, such as $\alpha$-RuCl$_3$. Not only in linear response (as in Refs.~[\onlinecite{bolens1},\onlinecite{bolens2}]), but as a tuning parameter that could shed light on whether a Kitaev spin liquid is present and potentially generate a new spin liquid with a Majorana-Fermi surface. Many questions remain, such as addressing the dynamics of this gapless liquid, the effect of temperature and the role of dilution and bond-disorder on this state. The answers to these questions have proven a rich source of insight in Kitaev's model. We hope that future work on liquids with Majorana-Fermi surfaces, as in the models presented in this work, can be similarly fruitful.

\begin{acknowledgements}
We thank A. Bolens, Y. B. Kim, P. A. McClarty, K. Penc and I. Sodemann  for helpful comments and discussions. This work was in part supported by Deutsche Forschungsgemeinschaft (DFG) under grant SFB 1143 and through the W\"urzburg-Dresden Cluster of Excellence on Complexity and Topology in Quantum Matter -- \textit{ct.qmat} (EXC 2147, project-id 39085490).
\end{acknowledgements}

\appendix

\section{$O(B^3)$ contributions to the effective Hamiltonian}

\begin{figure}[b]
    \centering
    \begin{tikzpicture}[baseline=-\R]
    \draw[fill=black!15!white] (0:\R) \foreach \x in {60,120,...,359} {-- (\x:\R)}-- cycle (90:\R);
    \draw[xshift=-1.5\R,fill=black!15!white,yshift=-0.8660254\R] (0:\R) 
    \foreach \x in {60,120,...,359} {-- (\x:\R)}-- cycle (90:\R);           
    \draw[yshift=-1.73205\R,fill=black!15!white] (0:\R) 
    \foreach \x in {60,120,...,359} {-- (\x:\R)}-- cycle (90:\R);
    \node[color=black,circle,draw,circle,inner sep=1.25pt,fill=white] at (-1.0\R,0.0) {};
    \node at (-1.3\R,0.4\R) {$\s^x_{i+x}$};
    \node[color=black,circle,draw,circle,inner sep=1.25pt,fill=white] at (-1.0\R,-2*0.8660254\R) {};
    \node at (-1.3\R,-2*0.8660254\R-0.5\R) {$\s^y_{i+y}$};
    \node[color=black,circle,draw,circle,inner sep=1.25pt,fill=white] at (0.5\R,-0.8660254\R) {};
    \node at (1.0\R,-0.8660254\R) {$\s^z_{i+z}$};
    \draw[black,very thick] (-0.5\R,-0.8660254\R) -- (0.5\R-0.07\R,-0.8660254\R);
    \draw[black,very thick] (-0.5\R,-0.8660254\R) -- (-1.0\R+0.05\R,0.0\R-0.05\R);
    \draw[black,very thick] (-0.5\R,-0.8660254\R) -- (-1.0\R+0.05\R,-2*0.8660254\R+0.05\R);
    \node at (-0.5\R+0.1\R,-0.8660254\R+0.25\R) {$i$};
    \end{tikzpicture}
    \caption{Illustration of a contribution to the $O(B^{3})$ part of the effective Hamiltonian, for $i\in A$. The relevant pieces of the magnetization operator are indicated by thick bonds.}
    \label{fig:third-order-mag}
\end{figure}
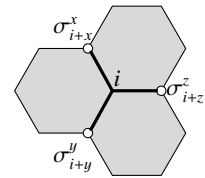

The most general term we can write at $O(B^{3})$ in our perturbation theory is given by:
\begin{equation*}
    -(g \mu_B)^{3} \sum_{\mu,\nu,\lambda}\sum_{i,j,k}  B_{\mu} B_{\nu} B_{\lambda}  \left[ P_0 \s^{\mu}_{i} \left(\frac{1-P_0}{\Delta}\right)  \s^{\nu}_{j} \left(\frac{1-P_0}{\Delta}\right) \s^{\lambda}_{k} P_0 \right],
\end{equation*}
where, as in the main text, we have used that the resolvent reduces to $R = (1-P_0)/\Delta$.
From the flux patterns generated by each piece (see Fig.~\ref{fig:third-order-mag}), we see that to obtain a combination of operators that leads us back to the zero state flux sector, one needs $\mu,\nu,\lambda$ to be a permutation of $x$, $y$, $z$. All of these permutations yield the same operator, and thus can be accounted for with an overall factor of $3!=6$. Noting that $P_0 \vec{M} P_0=0$ we can see that the only non-zero terms are
\begin{align*}
    &-\frac{6 h_{x}h_{y}h_{z}}{\Delta^{2}} \sum_{i,j,k} \left[ P_0 \s^{x}_{i} \s^{y}_{j} \s^{z}_{k} P_0 \right].
\end{align*}
There are two choices for $\{i,j,k\}$ which will give us non-zero contributions. One is given by $i=j=k\in \{A,B\}$
\begin{align*}
    &-\frac{6 h_{x}h_{y}h_{z}}{\Delta^{2}} \sum_{i} \left[ P_0 \s^{x}_{i} \s^{y}_{i} \s^{z}_{i} P_0 \right]=-\frac{6 h_{x}h_{y}h_{z}}{\Delta^{2}}(iN),
\end{align*}
which just gives an unimportant constant. The second non-zero contribution is generated by a configuration such as the one shown in Fig. ~\ref{fig:third-order-mag}.

To make this more explicit, write the $O(B^{3})$ correction (illustrated in Fig.~\ref{fig:third-order-mag}) as
\begin{align*}
    &-\frac{6 h_{x}h_{y}h_{z}}{\Delta^{2}} \Big\{\sum_{i\in A} \left[ P_0 \s^{x}_{i+x} \s^{y}_{i+y} \s^{z}_{i+z} P_0 \right] + \sum_{i\in B} \left[ P_0 \s^{x}_{i-x} \s^{y}_{i-y} \s^{z}_{i-z} P_0 \right]\Big\}.
\end{align*}
Finally, writing these operators in terms of the Majorana fermions we arrive at 
\begin{equation}
    -\frac{6 h_x h_y h_z}{\Delta^2}\sum_{{}^2\avg{ij}_{\alpha(\beta)\gamma}}
    \epsilon_{\alpha\beta\gamma}
    ic_i c_j,
\end{equation}
as stated in Sec.~\ref{sec:solution} [Eq.~(\ref{eq:three-spin-maj})].
\label{app:cubic}

\bibliography{draft}
\end{document}